\documentclass[12pt]{article}

\usepackage{epsfig}
\usepackage{amssymb}

\usepackage{graphics}
\usepackage{epsfig}
\usepackage{amssymb}
\usepackage{textcomp}
\usepackage{amsmath}
\newenvironment{Eqnarray*}{\arraycolsep 0.14em \begin{eqnarray*}}{\end{eqnarray*}}
\renewcommand{\theequation}{\mbox{\arabic{equation}}}
\newcounter{saveeqn}

\begin{document}
\begin{titlepage}
\begin{flushright}
        MZ-TH/05-24
\end{flushright}

\vskip 2.0in
\begin{center}
{\large {\bf Helicity Analysis of Semileptonic Hyperon Decays Including Lepton
Mass Effects}}
\end{center}
\begin{center}
\vskip 0.2in {\bf A.~Kadeer}, {\bf J.~G.~K\"orner} and {\bf U.~Moosbrugger}
\vskip 0.1in
{\it Institut f\"ur Physik, Johannes Gutenberg-Universit\"at,
D-55099 Mainz , Germany }\\

\vskip 0.2in {\bf Abstract}
\end{center}
\baselineskip=18pt 
\noindent

Using the helicity method we derive complete formulas for the joint 
angular decay distributions occurring in semileptonic
hyperon decays including lepton mass and polarization effects.
Compared to the traditional covariant calculation the helicity method
allows one to organize the calculation of the angular decay 
distributions in a very compact and efficient way. In the helicity method
the angular analysis is of cascade type, i.e. each decay in the decay 
chain is analyzed in the respective rest system of that 
particle. Such an approach is ideally suited as input for a
Monte Carlo event generation program.  As a specific example we take the decay 
$\Xi^0 \to \Sigma^+ + l^- + \bar{\nu}_l$ \,($l^-=e^-, \mu^-$) followed by the 
nonleptonic
decay $\Sigma^+ \to p + \pi^0$ for which we show a few examples of
decay distributions which are generated from a Monte Carlo program based on
the formulas presented in this paper. All the results of this paper are also
applicable to the semileptonic and nonleptonic decays of ground state charm and
bottom baryons, and to the decays of the top quark.
\end{titlepage}
\baselineskip=18pt


\section{Introduction \label{section-intro}}
Semileptonic hyperon decays have traditionally been analyzed in the rest frame of
the decaying parent hyperon using fully covariant methods based on either four-component 
Dirac spinor methods \cite{a59,h60,mmn68,rothe68,linke69,linke70,gk85} or on 
two-component Pauli spinor methods \cite{alles62,ww69,Bright:1999iy}. 
The latter method
is particularly well suited for an implementation of a zero-- \cite{alles62} or a near 
zero--recoil \cite{ww69,Bright:1999iy} approximation. In the present paper we employ 
helicity methods to analyze semileptonic 
hyperon decays. In the muonic mode it is quite important to incorporate lepton mass 
effects in the analysis since e.g. in the decay $\Xi^0 \to \Sigma^+ + l^- + \bar{\nu}_l$ 
the mass difference between the parent and daughter hyperon 
$M_{\Xi^0}- M_{\Sigma^+}=(1314.83 -1189.37)\,\rm{MeV}=125.46\,{\rm MeV}$ is 
comparable to the
muon mass $m_\mu=105.658 \,\rm{MeV}$.
The analysis proceeds in cascade fashion where every decay in the 
decay chain is analyzed in its respective rest frame. For the semileptonic decays
$\Xi^0 \to \Sigma^+  (\to p + \pi^0)+ W^-_{\rm{off-shell}} (\to l^- + \bar{\nu}_l)$
$(l=e^-,\mu^-)$
treated in this paper this means that the decay 
$\Xi^0 \to \Sigma^+ + W^-_{\rm{off-shell}} $ is analyzed in the $\Xi^0$ rest frame 
whereas the decays $\Sigma^+ \to p + \pi^0$ and 
$W^-_{\rm{off-shell}} \to l^- + \bar{\nu}_l$ are analyzed in the $\Sigma^+$ and
$W^-_{\rm{off-shell}}$ rest frames, respectively.
In this way one obtains exact decay formulas with no approximations which are 
quite compact since they can be written in a quasi--factorized form. 

Cascade--type analysis were quite popular some time ago in the strong interaction 
sector when analyzing the decay chains of the strong interaction baryonic and mesonic 
resonances (see e.g. \cite{kz68,ksz70,kz70,d71}). In the weak 
interaction sector cascade--type analysis were applied before to nonleptonic decays 
\cite{gk79,kp92,Korner:1991ap,bkkz93,Korner:1994nh,Shulga:2005zf,Ajaltouni:2004zu,Leitner:2004dt}, 
to semileptonic decays of heavy mesons and baryons 
\cite{Korner:1991ap,bkkz93,Korner:1994nh,Korner:1987kd,ks90,Hagiwara:1989gz,Hagiwara:1989cu,hmw90,Korner:1991ph,Bialas:1993pi,Ali:2002qc},
and to rare decays of heavy mesons 
\cite{Ali:2002qc,Faessler:2002ut} and
heavy baryons \cite{Aliev:2005np}. 
A new feature appears in semileptonic decays compared to nonleptonic decays when one
includes lepton
mass effects. In this case one has new interference contributions 
coming from the time-components of the 
vector and axial vector currents interfering with the usual three-vector
components of the currents (see e.g. \cite{bkkz93,ks90}).

The results for the angular decay distributions in the semileptonic decays of heavy 
baryons given e.g. in \cite{bkkz93,Korner:1991ph} can in fact be directly transcribed 
to the hyperon sector~\footnote{Our approach is similar in spirit to the approach of 
\cite{ft71} which also used a cascade--type helicity analysis to analyze the 
semileptonic decay of a polarized hyperon using the lepton side as a polarization 
analyzer.}. However, the presentation in \cite{bkkz93,Korner:1991ph} is
rather concise and concentrates on results for angular decay distributions and 
their analysis
rather than presenting the details of their derivations. In order to make the results 
more reproducible we decided to include the details of the derivations in this paper. 
This will enable the interested reader to e.g. convert the results of \cite{bkkz93}, which 
were derived for the $(l^+,\nu_{l})$ case, to the $(l^-,\bar{\nu}_l)$ case discussed in this
paper, or, to derive angular decay formula involving the decay $\Sigma^+ \to p + \gamma$ 
instead of the decay $\Sigma^+ \to p + \pi^0$ treated in this paper.
At the same time we decided to recalculate all relevant decay formulas in order to provide 
another independent check of their correctness. In this way we discovered one error 
in \cite{Korner:1991ph} and two errors in \cite{bkkz93}.



In the simulation of semileptonic hyperon decays including the
$\mu$--mode it is important to have a reliable and tested Monte Carlo (MC) program.
Since hyperons are produced with nonzero polarization the MC program 
should also include polarization effects of the decaying parent
hyperon. One of the motivations of starting this project was the fact that such a 
general purpose MC event generator has not been available up to now. Such an event
generator should prove to be quite useful in the analysis of the huge amount of data on
semileptonic hyperon decays that has been collected by the KTeV and the NA48 
collaborations. We wrote and tested 
such a MC event generator based on the formulas written down in this paper. The present
paper can be viewed as a documentation of the theoretical spin-kinematical input that 
goes into the MC program and, for the sake of reproducibility, the paper also 
describes how to derive the angular decay distributions entering the MC. 

Although we frequently refer to the specific semileptonic cascade decay
$\Xi^0 \to \Sigma^+ (\to p + \pi^0) + W^-_{\rm{off-shell}} (\to l^- + \bar{\nu}_l)$
the spin--kinematical analysis presented in this paper is quite general and can
be equally well applied to the semileptonic decays of heavy charm and bottom baryons,
and for that matter, also to the semileptonic decay of the top quark. 
In order to facilitate such further applications we have always included the 
necessary sign changes when going from the $(l^-,\bar{\nu}_l)$ case to the 
$(l^+,\nu_l)$ case as occurs e.g. in the semileptonic hyperon decay 
$\Sigma^+ \to \Lambda + e^+ + \nu_e$, in semileptonic $c \to s$ charm baryon decays
or in semileptonic top quark decays 
\cite{Fischer:1998gs,Fischer:2000kx,Fischer:2001gp,Do:2002ky}.
When sign changes are indicated the upper sign will always refer to the 
$(l^-,\bar{\nu}_l)$ case, which is the main concern of this paper, whereas the lower 
sign will refer to the $(l^+,\nu_l)$ case. We also mention that we have always
assumed that the amplitudes are relatively real and have therefore dropped
azimuthal correlation contributions coming from the imaginary parts. Put
in a different language this means that we have not considered $T$--odd
contributions in our angular analysis which could result from final state interaction
effects or from truly $CP$--violating effects. By keeping the imaginary parts in the
azimuthal correlation terms one can easily write down the relevant $T$--odd 
contributions if needed by using the formulas of this paper. This is discussed for a 
specific example in Appendix~\ref{appendix-T-odd}.

The paper is structured as follows. 
In Sec.~\ref{section-helicity-ampt}
we introduce the helicity amplitudes and
relate them to a standard set of invariant form factors. In order to estimate the size
of the helicity amplitudes for the $\Xi^0 \to \Sigma^+$ current--induced trasition
we provide some simple estimates for the invariant form factors and their
$q^2$--dependence which we shall refer to as the minimal form factor model. 
In Sec.~\ref{section-unpolarized} we derive the unpolarized decay rate written 
in terms of bilinear forms of the helicity amplitudes.
Sec.~\ref{section-numerics} contains some numerical results on branching ratios, 
the rate ratio $\Gamma(e)/\Gamma(\mu)$ and a lepton--side forward-backward asymmetry. 
In Sec.~\ref{section-single-spin} we discuss single spin polarization effects 
including spin--momentum correlation effects between the polarization of the parent 
baryon and the momenta of the decay products. 
Sec.~\ref{section-joint} treats momentum-momentum correlations between the 
momenta of the decay products in the cascade decay 
$\Xi^0 \to \Sigma^+ (\to p + \pi^0) + W^-_{\rm{off-shell}} (\to l^- + \bar{\nu}_l)$ 
for an unpolarized $\Xi^0$. 
In Sec.~\ref{section-montecarlo} we present a few sample distributions generated 
from the MC program written by us. 
Sec.~\ref{section-summary} contains our summary and our conclusions. 

We have collected some technical material in the appendices. 
In Appendix~\ref{appendix-2body-master} we recount
how the two-body decay of a polarized particle is treated in the helicity formalism. 
This two-body decay enters as a basic building block in our quasi-factorized master 
formulae in the main text which describe the various cascade--type
angular decay distributions presented in this paper. In Appendix~\ref{appendix-Wigner-D} 
we list explicit 
forms of the Wigner's $d^J$--function for $J=1/2$ and $J=1$, which are needed in the
present application. In Appendix~\ref{appendix-T-odd} we go through a specific example
and identify a specific $T$--odd term in the joint angular decay distribution
written down in Sec.~\ref{section-joint}. The example is easily generalized to other cases. 
In Appendix~\ref{appendix-5fold} 
we finally list the full five--fold angular decay distribution for the cascade decay
$\Xi^0 \to \Sigma^+ (\to p + \pi^0) + W^-_{\rm{off-shell}} (\to l^- + \bar{\nu}_l)$ for 
a polarized parent hyperon $\Xi^0$. The full five--fold angular decay distribution
reduces to the decay distributions listed in the main text after integration or after 
setting the relevant parameters to zero.   

\section{The helicity amplitudes \label{section-helicity-ampt}}

The momenta and masses in the semileptonic hyperon decays are denoted
by $B_1(p_1,M_1)\to B_2(p_2,M_2)+ l(p_l,m_l) + \nu_l(p_\nu,0)$. For the hadronic
transitions described by the helicity amplitudes it is not necessary to distinguish 
between the cases $(l^-,\bar{\nu}_l)$ and $(l^+,\nu_l)$.
The matrix elements of the vector and axial vector currents $J_\mu^{V,A}$ 
between the spin $1/2$ states are written as
\begin{align}
\label{covariant1}
M_\mu^V &= \langle B_2|J_\mu^V|B_1\rangle  = \bar{u}_2(p_2)\bigg[F_1^V(q^2)\gamma_\mu 
 + \frac{F_2^V(q^2)}{M_1}\sigma_{\mu \nu}q^\nu+ \frac{F_3^V(q^2)}{M_1}q_\mu \bigg] u_1(p_1)\;,\\
\label{covariant2}
M_\mu^A & =\langle B_2|J_\mu^A|B_1 \rangle = \bar{u}_2(p_2)\bigg[F_1^A(q^2)\gamma_\mu
+ \frac{F_2^A(q^2)}{M_1}\sigma_{\mu \nu}q^\nu  + \frac{F_3^A(q^2)}{M_1}q_\mu \bigg]\gamma_5 u_1(p_1)\;,
\end{align}
where $q=p_1-p_2$ is the four--momentum transfer. As in \cite{gk85} we take 
$\sigma_{\mu \nu}= \frac{1}{2}(\gamma_\mu \gamma_\nu - \gamma_\nu \gamma_\mu)$
and $\gamma_5=-\left(\begin{array}{cc}0 & 1 \\ 1 & 0
\end{array}
\right)$. The other $\gamma$ matrices are defined as in Bjorken-Drell.

Next we express the vector and axial vector helicity amplitudes 
$H_{\lambda_{2} \lambda_W}^{V,A}$ 
$(\lambda_{2}=\pm 1/2$; 
$\lambda_W=t,\pm1,0$)
in terms of the
invariant form factors, where $\lambda_W$ and $\lambda_{2}$ are the helicity components
of the $W_{\rm{off-shell}}$ and the daughter baryon, respectively. 
Since lepton mass effects are taken into account
in this paper we need to retain the time--component ``$t$'' of the four-currents
$J^{V,A}_\mu$. Concerning the transformation properties of the four components of the
currents one notes
that, in the rest frame of the $W_{\rm{off-shell}}$ ($\vec{q}=0$), the three 
space--components $\lambda_W=\pm1,0$ transform as $J=1$ whereas the time-component 
transforms as $J=0$. In this paper we use a short--hand notation
$\lambda_W=t,\pm1,0$ for $\lambda_W=0(J=0);\pm 1,0(J=1)$. Whenever we write
$\lambda_W=t$ this has to be understood as $\lambda_W=0(J=0)$.

One then needs to calculate the expressions  
\begin{equation}
\label{helamp1}
H^{V,A}_{\lambda_2 \lambda_W} =M_\mu^{V,A}(\lambda_2)
\bar{\epsilon}^{*\mu}(\lambda_W) \,\, .
\end{equation}
We do not explicitly denote the helicity ($m$--quantum number)
of the parent hyperon $\lambda_1$ in the 
helicity amplitudes since
$\lambda_1$ is fixed by the relation $\lambda_1=\lambda_2-\lambda_W$.
It is very important to detail the phase conventions when evaluating the expression
in Eq.~(\ref{helamp1}). This is because the angular decay distributions to be discussed later 
on contain interference contributions between different helicity amplitudes which 
depend on the relative signs of the helicity amplitudes. We shall work in the
rest frame of the parent baryon $B_1$ with the daughter baryon $B_2$ moving in the
positive $z$--direction. The baryon spinors are then given by \cite{ab66}
\begin{eqnarray}
\label{spinor}
\bar{u}_2(\pm{\textstyle \frac{1}{2}},p_2)&=& \sqrt{E_2+M_2}\left( \chi_\pm^\dagger,
\frac{\mp|\vec{p}_2|}{E_2+M_2}\chi_\pm^\dagger \right) \,,\nonumber \\
u_1(\pm{\textstyle \frac{1}{2}},p_1) &=& \sqrt{2M_1}
\left(\begin{array}{c}\chi_\pm \\ 0 \end{array}\right) \, ,
\end{eqnarray}
where $\chi_+=\left(\begin{array}{c}1 \\ 0 \end{array}\right)$ and 
$\chi_-=\left(\begin{array}{c}0 \\ 1 \end{array}\right)$ are the usual Pauli 
two-spinors. For the four 
polarization four-vectors of the currents we have \cite{ab66}
\begin{eqnarray}
\label{polvec}
\bar{\epsilon}^\mu(t) &=& \frac{1}{\sqrt{q^2}}\left(q_0;0,0,-p\right)\,, \nonumber \\
\bar{\epsilon}^\mu(\pm 1) &=& \frac{1}{\sqrt{2}}\left(0;\pm 1,-i,0\right) \,,\nonumber \\
\bar{\epsilon}^\mu(0) &=& \frac{1}{\sqrt{q^2}}\left(p;0,0,-q_0\right)\,, 
\end{eqnarray}
where $q^{\mu}=(q_{0};0,0,-p)$ is the momentum four--vector of the off--shell
gauge boson $W_{\rm{off-shell}}$ in the $B_{1}$ rest frame.
The energy of the off-shell $W$--boson $q_{0}$ and 
the magnitude of three--momentum of the daughter baryon $B_2$ (or the off-shell
$W$--boson) $p$
in the rest system of the parent baryon $B_1$ are given by  
\begin{align}
q_0 &=\frac{1}{2M_{1}}(M_1^2-M_2^2+q^2)\,,\\
p &=|\vec{p}_2|=\frac{1}{2M_{1}}\sqrt{Q_+Q_-}\,,
\end{align}
where
\begin{equation}
Q_\pm=(M_1 \pm M_2)^2 - q^2\,.
\end{equation}
The bar over the polarization four-vectors reminds one that the
$m$ quantum numbers of the currents are quantized along the negative $z$--direction.
They are obtained from the polarization four-vectors quantized along the
positive $z$--axis by a $180^\circ$ rotation around the $y$--axis (see \cite{ab66}).
Using the spinors in Eq.~(\ref{spinor}) and the polarization vectors Eq.~(\ref{polvec}) one obtains
following vector helicity amplitudes ($\lambda_1=\lambda_2-\lambda_W$)
\begin{eqnarray}
\label{vectorhel}
H_{\frac{1}{2}t}^V &=& \frac{\sqrt{Q_+}}{\sqrt{q^2}}\bigg(
(M_1-M_2) F_1^V +  q^2/M_1 F_3^V \bigg) \,,\nonumber \\
H_{\frac{1}{2}1}^V &=&\sqrt{2Q_-}\bigg(
- F_1^V - (M_1+M_2)/M_1 F_2^V \bigg) \nonumber \,,\\
H_{\frac{1}{2}0}^V &=& \frac{\sqrt{Q_-}}{\sqrt{q^2}}\bigg(
(M_1+M_2) F_1^V + q^2/M_1 F_2^V \bigg) \,,
\end{eqnarray}
and axial vector helicity amplitudes
\begin{eqnarray}
\label{axialhel}
H_{\frac{1}{2}t}^A &=& \frac{\sqrt{Q_-}}{\sqrt{q^2}}\bigg(
-(M_1+M_2) F_1^A +  q^2/M_1 F_3^A \bigg) \,,\nonumber \\
H_{\frac{1}{2}1}^A &=& \sqrt{2Q_+}\bigg(
 F_1^A - (M_1-M_2)/M_1 F_2^A \bigg) \; \nonumber \,,\\
H_{\frac{1}{2}0}^A &=& \frac{\sqrt{Q_+}}{\sqrt{q^2}}\bigg(
-(M_1-M_2) F_1^A + q^2/M_1 F_2^A \bigg)\,. 
\end{eqnarray}

From parity or from an explicit calculation one has
\begin{eqnarray}
H_{-\lambda_2, -\lambda_W}^{V} &=& H_{\lambda_2, \lambda_W}^{V}\, \nonumber\\
H_{-\lambda_2, -\lambda_W}^{A} &=& -H_{\lambda_2, \lambda_W}^{A}
\end{eqnarray}

When discussing the semileptonic transitions close to the zero recoil
point it is advantageous to
make use of the velocity transfer variable $\omega=v_1 \cdot v_2$
where $v^\mu=p^\mu/M$. The
velocity transfer variable can be expressed in terms of the usual momentum 
transfer variable $q^2$. One has
\begin{equation}  
\omega=(M_1^2+M_2^2-q^2)/(2M_1M_2)\,,
\end{equation}
or
\begin{equation}
q^2=(M_1-M_2)^2-2M_1M_2(\omega - 1)\,\, .
\end{equation}
The maximal and minimal values of the velocity transfer variable $\omega$ are 
$\omega_{\rm{max}}=(M_1^2+M_2^2-m_l^2)/(2M_1M_2)$
and $\omega_{\rm{min}}=1$, respectively. The minimal value $\omega_{\rm{min}}=1$ 
is referred to as the zero recoil point since this is the point where the recoiling 
daughter baryon has no three--momentum.
For the variables $Q_{\pm}$ defined in Sec.~\ref{section-helicity-ampt} one finds
$Q_{\pm}= 2 M_1 M_2 (\omega \pm 1)$ which gives $p=M_2\sqrt{\omega^2-1}$ where $p$
is the momentum of the recoiling daughter baryon.
The relevant expansion parameter close to the zero recoil point is thus
$\sqrt{\omega-1}$. For example, at the zero
recoil point only the helicity amplitudes $H_{\frac{1}{2}t}^V$ 
(allowed Fermi transition)
and $H_{\frac{1}{2}0}^A=-H_{\frac{1}{2}1}^A/\sqrt{2}$ (allowed Gamov--Teller
transition) survive. In the LS--coupling scheme with LS--amplitudes $T_{LS}$, 
these correspond to the $S$--wave transition amplitudes $T^V_{0\frac{1}{2}}$ and 
$T^A_{0\frac{1}{2}}$, respectively, where the orbital angular momentum $L$ is defined 
with respect to the relative orbital motion of the baryon $B_2$ and the $W^-_{\rm{off-shell}}$ 
in the rest frame of the baryon $B_1$. In the literature one can find very ingenious 
approximation formulae 
for various decay distributions and polarization observables which are based on a 
near zero recoil expansion \cite{ww69,Bright:1999iy}. These are usually referred to as 
effective theories of semileptonic hyperon decays. In this paper we shall, however, 
not discuss zero recoil or near zero recoil approximations but we always retain 
the full structure of the physical observables without any approximations. 

In order to get a feeling about the size of the helicity amplitudes we make a simple
minimal ansatz for the invariant amplitudes at zero momentum transfer using 
$SU(3)$ symmetry. The analysis is greatly simplified by the fact that the C.G. 
coefficients for the $(n \to p)$--transition are the same as those for the 
$(\Xi^0 \to \Sigma^+)$--transition. One thus has $F_1^V(0)=1$ and $F_1^A(0)=1.267$
where the vector form factor $F_1^V(0)$ is protected from first order symmetry
breaking effects by the Ademollo-Gatto theorem \cite{ag64, ag64-2}.
For the magnetic form factor $F_2^V(0)$ we take 
$F_2^V(0)=M_{\Xi^0}(\mu_p-\mu_n)/(2M_p)=2.6$
as in \cite{gk85}, 
where $\mu_{p}$ and $\mu_{n}$ are the anomalous magnetic moments of the proton and neutron.
The second class current contributions are
set to zero, i.e. we take $F_3^V(0)=F_2^A(0)=0$. Note that a first class quark current 
can in principle populate the second class form factors $F_3^V$ and $F_2^A$ when 
$M_{1}\neq M_{2}$. For example, in the covariant
spectator quark model calculation of \cite{Hussain:1990ai,Hussain:1990mh}
one finds $F_3^V(0)=(M_1-M_2)/(6M_2)=0.0176$ and 
$F_2^A(0)=0$. However, since we are not including $SU(3)$ symmetry breaking effects
for the other form factors we set $F_3^V$ and $F_2^A$ to zero for consistency reasons
since they vanish in the $SU(3)$ symmetry limit. For 
$F_3^A(0)$ we use the Goldberger-Treiman relation 
$F_3^A(0)=M_{\Xi^0}(M_{\Xi^0}+M_{\Sigma^+})F_1^A(0)/(m_{K^-})^2 = 17.1$ (see e.g. 
\cite{brene64}).

For the $q^2$--dependence of the invariant form factors we take a Veneziano--type ansatz
which has given a good description of the $q^2$--dependence of the electromagnetic form 
factors of the neutron and proton \cite{Korner:1976hv}. We write
\begin{align}
\label{venez}
F_i^{V,A}(q^2)&=F_i^{V,A}(0)\prod_{n=0}^{n_i}\frac{1}{1-\frac{q^2}{m_{V,A}^2+
n \alpha^{'-1}}} \approx F_i^{V,A}(0)\left(1 + q^2 \sum_{n=0}^{n_i}\frac{1}{m_{V,A}^2
+n \alpha^{'-1}}\right)\,.
\end{align}
For $F_1^V(q^2)$ and $F_2^V(q^2)$ we use $m_V=m_{K^*(892)}=0.892$ GeV which is the 
lowest lying strange vector meson with $J^P=1^-$ quantum numbers. Correspondingly we use
$m_A=m_{K^*(1.270)}=1.273\, {\rm GeV}$ ($J^P=1^+$) for $F_1^A(q^2)$ and $m_A=m_K=0.494\, {\rm GeV}$
($J^P=0^-$) for $F_3^A(q^2)$, respectively. The slope of the Regge trajectory is taken 
as $\alpha'=0.9\, {\rm GeV}^{-2}$. The number of poles in Eq.~(\ref{venez}) is determined by 
the large $q^2$ power counting laws. One thus has $n_{1}=1$ and $n_{2,3}=2$. For the 
slopes of the form factors we thus have 1.781, 2.113, 0.983 and 5.241 $\rm{GeV}^{-2}$ 
for $F_{1}^V, F_{2}^{V}, F_1^A$ and $F_3^A$, 
respectively. These values correspond to (``charge'') radii $ \langle r^{2}\rangle $ of
0.416, 0.494, 0.230 and 1.224 ${\rm fm}^{2}$, respectively, for the four above form 
factors. The (``charge'') radii of the $F_{1}^V$ and $F_1^A$ form factors are close to 
the radii
calculated in \cite{faessler08} using chiral input and a relativistic constituent
quark model.  
The $q^2$--dependence of the form factors introduces only small effects 
since the range of $q^2$ is so small for the $\Xi^0 \to \Sigma^+$ transitions. For the 
largest $q^2$--value $q^{2}_{\rm{max}}=(M_{1}-M_{2})^{2}$, the form factors have only 
increased by $2.8\%\,(F_{1}^V)$, $3.3\%\,(F_{2}^V)$, $1.5\%\,(F_1^A)$ and 
$8.2\%\,(F_3^A)$ from their $q^2=0$ values.

\begin{figure}[t]\begin{center}
\epsfig{figure=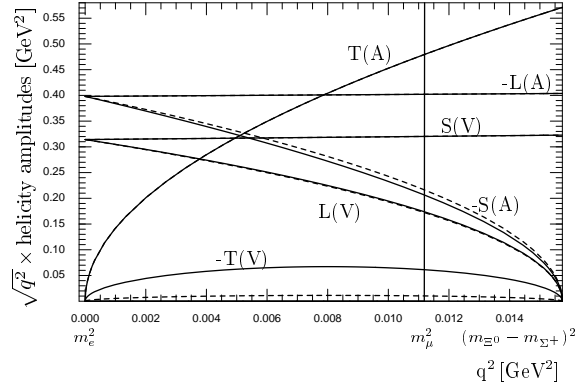, width=7.5cm}
\caption{\label{helamp} \small \sf
The $q^2$--dependence of the six independent helicity amplitudes 
$S(V,A):=\sqrt{q^2}H_{\frac{1}{2}t}^{V,A},
\,T(V,A):=\sqrt{q^2}H_{\frac{1}{2}1}^{V,A}$ and $L(V,A):=\sqrt{q^2}
H_{\frac{1}{2}0}^{V,A}$ multiplied by 
$\sqrt{q^2}$ for the $e$-mode (full range) and for the $\mu$--mode
(to the right of the vertical line $q^2 = m_\mu^2$). We also display the corresponding
helicity amplitudes with the contributions of $F_{2}^{V}(q^2)$ and $F_{3}^{A}(q^2)$ 
switched off (dashed lines).  }
\end{center}\end{figure}

Based on these estimates for the invariant form factors we show in Fig.~\ref{helamp} a 
plot of the $q^2$--dependence of the six helicity amplitudes.
For easier comparability we have plotted the quantities 
$\sqrt{q^2}H_{\lambda_2\lambda_W}^{V,A}$. Close to the lower boundary $q^2=m_e^2$ the 
longitudinal and scalar helicity amplitudes dominate, with
$H_{\frac{1}{2}0}^V \approx H_{\frac{1}{2}t}^V$ and 
$H_{\frac{1}{2}0}^A \approx H_{\frac{1}{2}t}^A$. Close to the upper boundary at the zero
recoil point $q^2=(M_1-M_2)^2$, which is the relevant region for the $\mu$--mode, the 
orbital $S$--wave contributions $H_{\frac{1}{2}t}^V$ and $H_{\frac{1}{2}1}^A=
-\sqrt{2} H_{\frac{1}{2}0}^A$ are the dominant amplitudes with 
$H_{\frac{1}{2}t}^V=-(F_{1}^{V}/F_{1}^{A})H_{\frac{1}{2}0}^A$ when 
$F_{3}^{V}=F_{2}^{A}=0$.
In Fig.~\ref{helamp} we also show plots of the helicity amplitudes where the 
contributions of 
the form factors $F_{2}^{V}(q^2)$ and $F_{3}^{A}(q^2)$ have been switched off.
The difference is not discernible at the scale of the plot for the helicity
form factors $H_{\frac{1}{2}0}^{V}$, $H_{\frac{1}{2}t}^{V}$, $H_{\frac{1}{2}1}^{A}$
and $H_{\frac{1}{2}0}^{A}$ and is small for $H_{\frac{1}{2}t}^{A}$. The difference is, 
however, sizeable for $H_{\frac{1}{2}1}^{V}$. This can be understood from 
Eq.~(\ref{vectorhel}) which shows that, even though $H_{\frac{1}{2}1}^{V}$ is  
small because it is proportional to $\sqrt{Q_{-}}$ and thus of overall order
$O((M_{1}-M_{2})/M_{1})$, the relative contribution 
of $F_{2}^{V}$ to $H_{\frac{1}{2}1}^{V}$ is not suppressed by a factor of
$O((M_{1}-M_{2})/M_{1})$ 
as it is in the other helicity form factors. Thus a measurement of the helicity
form factor $H_{\frac{1}{2}1}^{V}$ would be ideally suited to determine the strength
of $F_{2}^{V}$. In fact, in  Sec.~\ref{section-numerics} we shall discuss a 
forward-backward asymmetry measure which is proportional to $H_{\frac{1}{2}1}^{V}$
and is thus well suited for a measurement of $F_{2}^{V}$. 

We caution the reader that our ansatz for the form factors is only meant to
implement the gross features of the dynamics of the semileptonic 
hyperon decays $\Xi^0 \to \Sigma^+ + l^- + \bar{\nu}_l$ which will
eventually be superseded by the results of a careful analysis of the decay data.
We shall nevertheless use the above minimal model for the $\Xi^0 \to \Sigma^+$ form 
factors to calculate branching rates, the 
rate ratio $\Gamma(e)/\Gamma(\mu)$ and a forward-backward asymmetry in 
Sec.~\ref{section-numerics}, the longitudinal and transverse 
polarizations of the daughter baryon and the lepton in 
Sec.~\ref{section-single-spin} and a mean azimuthal correlation
parameter in Sec.~\ref{section-joint} for this decay. 
In Sec.~\ref{section-montecarlo} we also 
show some Monte Carlo plots
which are again based on the minimal form factor model.

In the subsequent sections the rate and the angular decay distributions will mostly be 
written in terms of bilinear products
of the sum of the vector and axial vector helicity amplitudes
\begin{equation}
\label{vplusa-helamp}
H_{\lambda_2 \lambda_W} = H_{\lambda_2 \lambda_W}^V + H_{\lambda_2\lambda_W}^A\,,
\end{equation}
since it is this combination which appears naturally in the master formulas describing
the rate and the various decay distributions.
One can of course rewrite the rate and the decay distributions in terms of bilinear 
products of the vector and axial vector helicity amplitudes 
$H_{\lambda_2 \lambda_W}^{V}$ and $H_{\lambda_2 \lambda_W}^{A}$. For the decay 
distributions this can be quite illuminating if one wishes to identify the overall 
parity nature of the observables that multiply the angular terms in the angular decay 
distributions.

\section{Unpolarized decay rate} \label{section-unpolarized}
The differential decay rate is given by (see e.g. \cite{ks90})
\begin{equation}
\label{diffrate2}
\frac{d\Gamma}{dq^2 dE_l} = \frac{G_F^2}{(2\pi)^3}
|V_{us}|^2 \frac{1}{8M_1^2} L_{\mu\nu}H^{\mu\nu}\;,
\end{equation}
where $L_{\mu\nu}$ is the usual lepton tensor ($\varepsilon_{0123}=+1$)
\begin{equation}
\label{ltensor}
L^{\mu \nu}= p_l^\mu p_\nu^\nu + p_l^\nu p_\nu^\mu - \frac{q^2-m_l^2}{2}g^{\mu\nu}
\pm i \varepsilon^{\mu \nu \alpha \beta}p_{l,\alpha} p_{\nu,\beta}\,\, . 
\end{equation}
As stated in the introduction the upper sign refers to the $(l^-,\bar{\nu}_l)$ case, 
whereas the lower sign refers to the $(l^+,\nu_l)$ case.

The hadron tensor $H_{\mu\nu}$ is given by the tensor product of the vector and
axial vector matrix elements defined in Eqs.~(\ref{covariant1}) and (\ref{covariant2}),
cf.
\begin{equation}
H_{\mu\nu}=(M^V+M^A)_\mu\,\,(M^V+M^A)_\nu^\dagger \,\, .  
\end{equation}
Eq.~(\ref{diffrate2}) shows
that $L_{\mu\nu}H^{\mu\nu}$ determines the
dynamical weight function in the $(q^2,E_l)$ Dalitz plot (see e.g. the discussion in
\cite{kajantie73}). In a Monte Carlo
generator one would thus have to generate events according to the weight
$L_{\mu\nu}H^{\mu\nu}$ in the $(q^2,E_l)$ Dalitz plot.

The differential $q^2$--distribution can be obtained from
Eq.~(\ref{diffrate2}) by $E_l$--integration, where the limits are given by (see e.g. \cite{ks90})
\begin{equation}
\label{el-limits}
E_l^\pm = \frac{1}{2q^2}\Big(q_0(q^2+m_l^2) \pm p(q^2-m_l^2)\Big).
\end{equation}

Finally, in order to get the total rate one has to integrate over $q^2$ in the limits
$m_l^2 \leqslant q^2 \leqslant (M_1-M_2)^2$.
 
On reversing the order of integrations, the differential lepton energy distribution can be 
obtained from
Eq.~(\ref{diffrate2}) by $q^2$--integration. The relevant integration limits can be obtained
from the inverse of Eq.~(\ref{el-limits}). One obtains (see e.g. \cite{ks90})
\begin{equation}
\label{q2-bound}
q^2_{\pm} = \frac{1}{a}(b \pm \sqrt{b^2-ac})\,\, ,
\end{equation} 
where 
\begin{eqnarray}
a &=& M_1^2 + m_l^2 - 2M_1 E_l \,,\nonumber  \\
b &=& M_1E_l(M_1^2-M_2^2 + m_l^2 - 2M_1E_l) + m_l^2 M_2^2\,, \nonumber \\
c &=& m_l^2\Big( (M_1^2-M_2^2)^2 + m_l^2M_1^2 - (M_1^2-M_2^2)2M_1E_l\Big)\, .
\nonumber
\end{eqnarray}
Using $E_l({\rm max}):= E_{{\rm max}}=(M_1^2-M_2^2+m_l^2)/(2M_1)$ and
$E_l({\rm min}):= E_{{\rm min}}= m_l$ one can simplify Eq.~(\ref{q2-bound}) to write
\begin{align}
q^2_{\pm} &=\frac{2M_1^2}{2M_1(E_{\rm{max}}-E_l)+M_2^2}
\bigg((E_{\rm{max}}-E_l)
 \Big(E_l \pm \sqrt{E_l^2-E_{\rm{min}}^2}\,\,\Big) + \frac{m_l^2M_2^2}{2M_1^2} \bigg).
\end{align}
Finally, in order to get the total rate, one has to integrate over the lepton 
energy in the limits $m_l \leqslant E_l \leqslant (M_1^2 - M_2^2 + m_l^2)/2M_1 $.

As it turns out the two--dimensional integration becomes much simpler if one 
considers the
two--fold differential rate with respect to the variables $q^2$ and $\cos\theta$ instead,
where $\theta$ is the polar angle of the lepton in the $(l,\nu_l)$ c.m.
system relative to the momentum direction of the $W_{\rm{off-shell}}$. $E_l$ and
$\cos\theta$ are related by (see e.g. \cite{ks90})
\begin{equation}
\label{costheta}
\cos \theta =\frac{2q^2E_l- q_0(q^2+m_l^2)}{p (q^2-m_l^2)}\,.
\end{equation}
Differentiating Eq.~(\ref{costheta}) one has
\begin{equation}
\frac{d \cos\theta}{dE_l} = \frac{2q^2}{p(q^2-m_l^2)} \,,
\end{equation}
which leads to the differential decay distribution 
\begin{equation}
\label{diffrate3}
\frac{d\Gamma}{dq^2 d\cos\theta} =  \frac{G_F^2}{(2\pi)^3}
|V_{us}|^2 \frac{(q^2-m_l^2)p}{16 M_1^2q^2} L_{\mu\nu}H^{\mu\nu}\;.
\end{equation}
It is clear from comparing Eqs.~(\ref{diffrate2}) and (\ref{diffrate3}) that, when
writing a Monte Carlo program, one should {\it not} generate events in the ($q^2,\cos\theta$) 
Dalitz plot according to the weight $L_{\mu\nu}H^{\mu\nu}$.

The $\cos\theta$ dependence of $L_{\mu\nu}H^{\mu\nu}$ can be easily worked out by 
following the methods described in \cite{ks90} which is based on the completeness relation
for the polarization four--vectors
\begin{equation}
\label{completeness}
 \sum_{m,m'=t,\pm,0}\bar{\epsilon}^{\mu }(m)\bar{\epsilon}^{* \nu }(m') g_{mm'} = 
g^{\mu \nu} \, .
\end{equation} 
\noindent The tensor $ g_{mm'} = \mbox{diag} (+,-,-,-) $ is the spherical representation
of the metric tensor where the components are ordered in the sequence $m,m'=t,\pm1,0$. 
One can then rewrite the contraction of the lepton and hadron tensors 
$L_{\mu\nu}H^{\mu\nu}$ as

\begin{align}
\label{lorentztrick}
L_{\mu\nu}H^{\mu\nu} & = L^{\mu'\nu'}g_{\mu'\mu}g_{\nu'\nu}H^{\mu\nu}
=\sum_{m,m',n,n'} L^{\mu'\nu'}\bar{\epsilon}_{\mu'}(m)\bar{\epsilon}_{\mu}^*(m')g_{mm'}
\bar{\epsilon}_{\nu'}^*(n)\bar{\epsilon}_{\nu}(n')g_{nn'}H^{\mu\nu}
\nonumber \\
&=\sum_{m,m',n,n'}
\bigg( L^{\mu'\nu'}\bar{\epsilon}_{\mu'}(m)\bar{\epsilon}_{\nu'}^*(n) \bigg)
 \bigg(H^{\mu\nu}\bar{\epsilon}_{\mu}^*(m')\bar{\epsilon}_{\nu}(n') \bigg) 
\,g_{mm'} g_{nn'} \,\, .
\end{align}
We shall refer to the second and third lines of Eq.~(\ref{lorentztrick}) as the semi-covariant 
representation of the angular decay distribution.

One has to remember that Eq.~(\ref{lorentztrick}) refers to the differential rate of
the decay of an unpolarized parent hyperon into a daughter baryon whose spin 
is not observed. This means that one has to take into account the additional conditions 
$\lambda_1=\lambda'_1$ and $\lambda_2=\lambda'_2$ in Eq.~(\ref{lorentztrick}). Angular
momentum conservation then implies that not all index pairs $m=m'$ and $n=n'$ in 
Eq.~(\ref{lorentztrick}) can be realized. Taking angular momentum conservation
into account one has diagonal contributions $m=m'=n=n'=t,\pm1,0$ as well as nondiagonal 
contributions with $m=m'=t$ and $n=n'=0$ and vice versa.

The point of writing $L_{\mu\nu}H^{\mu\nu}$ in the factorized form of 
Eq.~(\ref{lorentztrick}) is that each of the two factors in the second line of 
Eq.~(\ref{lorentztrick}) is Lorentz invariant and can thus
be evaluated in different Lorentz frames. The leptonic part will be evaluated
in the $(l,\nu_l)$ center--of--mass (c.m.) frame 
(or $W_{\rm{off-shell}}$--rest frame) bringing in the
decay angle $\theta$, whereas the hadronic
part will be evaluated in the $\Xi^-$ rest frame bringing in the helicity amplitudes
defined in Sec.~\ref{section-helicity-ampt}. 

Turning to the $(l,\nu_l)$ c.m system the lepton momenta in the $(x,y,z)$--system 
read (see Fig.~\ref{lepton-side})
\begin{eqnarray}
p_l^\mu &=& (E_l\,;p_l\sin\theta,0,-p_l\cos\theta) \,,\\
p_\nu^\mu &=& p_l\,(1,-\sin\theta,0,\cos\theta) \nonumber \, .
\end{eqnarray}
The angle $\theta$ is always measured with respect to the direction of the lepton $l$, regardless
of whether we are dealing with the $(l^-,\bar{\nu}_l)$ or the $(l^+,\nu_l)$ case.
Since the orientation in the $(x,y)$--plane need not be specified in the present problem
we have chosen the lepton momenta to lie in the $(x,z)$--plane.
$E_l$ and $p_l$ are the energy and the magnitude of the three-momentum of the
charged lepton in the $(l,\nu_l)$ c.m system, respectively, given by 
$E_l=(q^2+m_l^2)/2\sqrt{q^2}$ and $p_l=(q^2-m_l^2)/2\sqrt{q^2}$. The longitudinal
and time-component polarization four--vectors take the form 
$\bar{\epsilon}^\mu(0)=(0;0,0,-1)$
and $\bar{\epsilon}^\mu(t)=(1;0,0,0)$ whereas the transverse parts are unchanged from
Eq.~(\ref{polvec}). Using the explicit form of the lepton tensor Eq.~(\ref{ltensor}) it is 
then not difficult to evaluate Eq.~(\ref{lorentztrick}) in terms of
the helicity amplitudes $H_{\lambda_2\lambda_W}$ of Sec.~\ref{section-helicity-ampt}. 
One obtains

\begin{eqnarray}
\label{lepton-hadron1}
L_{\mu\nu}H^{\mu\nu}&=&\frac{2}{3}(q^2-m_l^2) \;
\bigg[  \frac{3}{8}(1\mp\cos \theta)^2
|H_{\frac{1}{2}1}|^2 + \frac{3}{8}(1\pm\cos \theta)^2 |H_{-\frac{1}{2}-1}|^2 \nonumber \\
&&+\; \frac{3}{4} \sin^2 \theta (|H_{\frac{1}{2}0}|^2 + |H_{-\frac{1}{2}0}|^2) \nonumber
 \\
&&+\; \frac{m_l^2}{2q^2} \;
\Big\{
\frac{3}{2}(|H_{\frac{1}{2}t}|^2 + |H_{-\frac{1}{2}t}|^2)
+\frac{3}{4}(|H_{\frac{1}{2}1}|^2 + |H_{-\frac{1}{2}-1}|^2) \sin^2\theta \nonumber \\
&&+\; \frac{3}{2}\cos^2 \theta(|H_{\frac{1}{2}0}|^2 + |H_{-\frac{1}{2}0}|^2)
 -3\cos \theta (H_{\frac{1}{2}t}H_{\frac{1}{2}0} 
+ H_{-\frac{1}{2}t}H_{-\frac{1}{2}0}) \Big\} \;\bigg]\;, 
\end{eqnarray}
where the $H_{\lambda_2 \lambda_W}$ are the sums of the corresponding vector and
axial vector helicity amplitudes defined in Eq.~(\ref{vplusa-helamp}). 
We mention that the helicity flip factor $m_l/2q^2$ does not give rise to a
singularity since $q^2 \ge m_l^2$.

By explicit verification, or by hindsight, one can show that Eq.~(\ref{lorentztrick}) can 
be written very compactly in terms of 
Wigner's $d^J$--functions. One has what we shall refer to as our first master formula
\begin{align}
\label{lepton-hadron2}
L_{\mu\nu}H^{\mu\nu}=&\frac{1}{8} \sum_{\lambda_l,\lambda_2,\lambda_W,\lambda'_W,J,J'}
(-1)^{J+J'} |h^l_{\lambda_l\lambda_\nu=\pm\frac{1}{2}}|^2 \times\nonumber \\
&\times \delta_{\lambda_2 - \lambda_W,\lambda_2 - \lambda'_W} 
d_{\lambda_W,\lambda_l \mp \frac{1}{2}}^J (\theta)
d_{\lambda'_W,\lambda_l\mp\frac{1}{2}}^{J'} (\theta)
H_{\lambda_2 \lambda_W}H_{\lambda_2 \lambda'_W}^\ast \;.
\end{align}
Except for the phase factor $(-1)^{J+J'}$ the master formula can in fact be derived by 
repeated application of the basic two--body decay formula in 
Appendix~\ref{appendix-2body-master}.
The Kronecker $\delta$-function $\delta_{\lambda_2 - \lambda_W,\lambda_2 - \lambda'_W}$ 
in Eq.~(\ref{lepton-hadron2}) expresses the fact that we are
dealing with the decay of an unpolarized parent hyperon.
One has to remember that $\lambda_W=0$ and $\lambda_W=t$ both refer to the helicity
projection 0 (see Sec.~\ref{section-helicity-ampt}). Therefore 
there are nondiagonal interference contributions between $J=1,\lambda_W=0$ and
$J=0,\lambda_W=t$ because they are allowed by the angular momentum
conservation condition $\lambda_2-\lambda_W=\lambda_2-\lambda'_W$ implying 
$\lambda_W=\lambda'_W$. The interference
contributions carry an extra minus sign as can be seen from the phase factor 
$(-1)^{J+J'}$ in Eq.~(\ref{lepton-hadron2}). The phase factor $(-1)^{J+J'}$ comes in 
because of the pseudo--Euclidean nature of the spherical metric tensor $g_{mm'}$
defined after Eq.~(\ref{completeness}). 

The sign change in the first line of Eq.~(\ref{lepton-hadron1}) going from the
$(l^-,\bar{\nu}_l)$ to the $(l^+,\nu_l)$ case can now be seen to result from the
products of the relevant elements of the Wigner's $d^1$--functions. For example,
for $\lambda_2=1/2$, $\lambda_W=1$ the nonflip contributions 
$(\lambda_l=-\lambda_\nu=\mp 1/2)$ are proportional to 
$(d^1_{1,\mp1})^2=({\textstyle \frac{1}{2}}(1\mp\cos\theta))^2$. There are no 
corresponding sign changes in the other lines of Eq.~(\ref{lepton-hadron1}).
 
The $h_{\lambda_l\lambda_\nu}$ are the helicity amplitudes of the
final lepton pair in the $(l,\nu)$ c.m. system. For example, for the $(l^-,\bar{\nu})$ 
case with $\vec{p}_{l^-}$ along the positive $z$-axis, they can be worked out by using 
Eq.~(\ref{spinor}), the negative energy spinor of 
the massless antineutrino with helicity $\lambda_{\bar{\nu}}=\frac{1}{2}$ given by
\begin{equation}
v_{\bar{\nu}}({\textstyle \frac{1}{2}}) = \sqrt{E_\nu}
\left(\begin{array}{c}\chi_+ \\-\chi_+  \end{array}\right) \, ,
\end{equation}
and the SM form of the lepton current ($\lambda_W = \lambda_{l^-} - \lambda_{\bar{\nu}}$) 
\begin{equation}
\label{lephel1}
h_{\lambda_{l^-}=\mp\frac{1}{2},\lambda_{\bar{\nu}}=
\frac{1}{2}}=\bar{u}_{l^-}(\mp{\textstyle \frac{1}{2}})\gamma^\mu(1+\gamma_5)v_{\bar{\nu}}
({\textstyle \frac{1}{2}})
\left\{{\epsilon_\mu (-1) \atop \epsilon_\mu (t),\epsilon_\mu (0)} \right\} \, .
\end{equation}
We shall refer to the upper case $\lambda_{l^-}=-\frac{1}{2}$ as the nonflip transition
and to the lower case $\lambda_{l^-}=\frac{1}{2}$ as the flip transition.
Note the unconventional form of the SM lepton current which is due to the $\gamma_5$ 
definition in Sec.~\ref{section-helicity-ampt}. The polarization four--vectors are given by 
$\epsilon^\mu(t)=(1;0,0,0)$, $\epsilon^\mu(0)=(0;0,0,1)$ and
$\epsilon^\mu(\pm 1)=(0;\mp1,-i,0)/\sqrt{2}$. The flip contribution is identical for
$\lambda_W=t$ and $\lambda_W=0$. 
A similar expression can be written down for the case $(l^+,\nu_l)$ which we shall 
not work out in explicit form. For the moduli squared of the helicity amplitudes one 
finally obtains 
\begin{alignat}{2}
\label{lephel2}
\rm{nonflip} \,\,(\lambda_W=\mp1): 
 |h_{\lambda_l=\mp\frac{1}{2},\lambda_\nu=\pm\frac{1}{2}}|^2
&= 8(q^2-m_l^2) \;, \\ \label{lephel3}
\rm{flip} \,\,(\lambda_W=t,0):
 |h_{\lambda_l=\pm \frac{1}{2},\lambda_\nu=\pm \frac{1}{2}}|^2
&= 8 \frac{m_l^2}{2q^2}(q^2-m_l^2)\;.
\end{alignat}
The notation in Eqs.(\ref{lephel2}) and (\ref{lephel3}) (and elsewhere in the paper) 
is such that the upper and
lower signs refer to the configurations $(l^{-},\bar{\nu}_{l})$ and 
$(l^{+},\nu_{l})$, respectively.
In Eq.~(\ref{lepton-hadron2}) the sum over $J,J'$ runs over $0$ and $1$ and the index
$\lambda_W,\lambda'_W$ runs over the four components $t,\pm1,0$. As remarked on before
one has to remember to include the interference contribution from $(J=0;\lambda_W=t)$ and 
$(J=1;\lambda_W=0)$ giving an extra minus sign. 
The matrix $d^1_{mm'}$, finally, is Wigner's $d^1$--function 
($d^0_{mm'}=1$ for $m,m'=t$) listed in Appendix B.

The form Eq.~(\ref{lepton-hadron2}) readily affords a physical interpretation. 
$H_{\lambda_2 \lambda_W}H_{\lambda_2 \lambda'_W}^\ast$ determines the density matrix of
the $W_{\rm{off-shell}}$ (which happens to be block-diagonal in the present application).
The density matrix is then ``rotated'' into the direction of the lepton in the
$(l,\nu_l)$ c.m. system with the help of the $d^1$--functions whence the squared 
helicity amplitudes $|h_{\lambda_l \lambda_\nu}|^2$ determine the helicity
dependent rates into the lepton pair. 

Performing the sum in Eq.~(\ref{lepton-hadron2}) $(\lambda_l=\pm 1/2;\lambda_W=t,\pm1,0;
J=0,1;J'=0,1;\lambda_2=\pm 1/2)$ one recalculates Eq.~(\ref{lepton-hadron1}).
Note that the flip contribution proportional to $m_l^2/2q^2$ and 
nonflip contributions 
are clearly separated in Eq.~(\ref{lepton-hadron1}). 
This separation facilitates the determination of the longitudinal polarization 
of the lepton to be discussed in Sec.~\ref{section-single-spin}. 

The differential rate $d\Gamma/dq^2$ is obtained from Eqs.~(\ref{diffrate3})
and (\ref{lepton-hadron1})
by $\cos\theta$--integration which, in a sense, is trivial. One obtains
\begin{align}
\label{diffrateq^2}
\frac{d\Gamma}{dq^2} =&\frac{1}{3} \frac{G_F^2}{(2\pi)^3}
|V_{us}|^2 \frac{(q^2-m_l^2)^2 p}{8M_1^2q^2}\bigg[  
|H_{\frac{1}{2}1}|^2 + |H_{-\frac{1}{2}-1}|^2 +
  |H_{\frac{1}{2}0}|^2 + |H_{-\frac{1}{2}0}|^2 \nonumber \\
&+\; \frac{m_l^2}{2q^2} \;
\Big\{
3\Big(\,|H_{\frac{1}{2}t}|^2 + |H_{-\frac{1}{2}t}|^2\,\Big) 
+|H_{\frac{1}{2}1}|^2  + |H_{-\frac{1}{2}-1}|^2  
+\; |H_{\frac{1}{2}0}|^2 + |H_{-\frac{1}{2}0}|^2
\Big\} \;\bigg]\;.
\end{align}
The remaining $q^{2}$--integration ($m_{l}^{2} \leqslant q^{2} \leqslant (M_{1}-M_{2})^{2}$) has
to be done numerically because of the nontrivial $q^{2}$--dependence of
the invariant form factors.
 
A check on Eq.~(\ref{diffrateq^2}) is afforded by recalculating the Standard
Model (SM) formula for semileptonic free quark decay $q_{1}\to q_{2}+ l + \nu$ 
setting $F_1^{V}(q^{2})=F_1^{A}(q^{2})=1$ and $F_{2,3}^{V,A}=0$ in 
Eq.~(\ref{diffrateq^2}). 
One obtains
\begin{align}
\label{smdiffrate}
\frac{d\Gamma^{SM}}{d\hat{q}^2} &= 
\Gamma_0 \frac{(\hat{q}^2-\eta^2)^2}{\hat{q}^4}
4 \hat{p} \bigg(-2\hat{q}^4+\hat{q}^2(1+\rho^2)+(1-\rho^2)^2\nonumber\\
& \quad +\frac{\eta^2}{2\hat{q}^2} \Big\{-2\hat{q}^4-2\hat{q}^2(1+\rho^2)
+4(1-\rho^2)^2 \Big\} \bigg)\, ,
\end{align}  
where we have introduced scaled variables according to $\hat{p}=p/M_1$, 
$\hat{q}^2=q^2/M_1^2$, $\rho^2=M_2^2/M_1^2$ and $\eta^2=m_l^2/M_1^2$, and where
$\hat{p}=\frac{1}{2}(1+\rho^4+\hat{q}^4-2\rho^2-2\hat{q}^2-2\rho^2\hat{q}^2)^{1/2}$
is the scaled magnitude of the daughter baryon's three momentum in the rest 
frame of the parent baryon. 
Also we have introduced the Born term rate
\begin{equation}
\Gamma_0=\frac{G_F^2|V_{us}|^2 M_1^5}{192 \pi^3}\,,
\end{equation}
which represents the Standard Model decay of a massive parent fermion into
three massless fermions, i.e. $M_1\neq 0$ and $M_2,m_l,m_\nu=0$. 
The result Eq.~(\ref{smdiffrate}) agrees with the SM result given e.g. in \cite{bkp88}.

The $\hat{q}^2$--integration of Eq.~(\ref{smdiffrate}) can now be done analytically. One 
obtains
$\big(\eta^2 \leqslant \hat{q}^2 \leqslant (1-\rho)^2\big)$
\begin{align}
\label{smrate}
\Gamma^{SM} =& \Gamma_0 \bigg[ L\Big(\frac{1}{2}-7\rho^2-7\rho^4
+\rho^6+6 \eta^2\rho^2-7\eta^4\rho^2\Big) \nonumber\\
&-24\rho^4(1-\eta^4)\ln\Big(\frac{1+\rho^2-\eta^2-L}{2\rho}
\Big)\; +(\rho \leftrightarrow \eta) \bigg] \,\, ,
\end{align}
where $L=(1+\rho^4+\eta^4-2\rho^2-2\eta^2-2\rho^2 \eta^2)^{1/2}$.
Eq.~(\ref{smrate}) again agrees with the result given in e.g. \cite{bkp88}.
The symmetrization $\rho \leftrightarrow \eta$ in Eq.~(\ref{smrate})  
must be done for both the logarithmic and 
nonlogarithmic terms. The symmetry of the rate
expression in Eq.~(\ref{smrate}) under the exchange $(\rho \leftrightarrow \eta)$ 
reflects the simple Fierz property of the SM $(V-A)$ coupling. We mention
that a less symmetric form of Eq.~(\ref{smrate}) has been written down in
\cite{cpt82}. 

We conclude this section with a comment on the relative merits of the two equivalent 
decay formulas in Eqs.~(\ref{lorentztrick}) and (\ref{lepton-hadron2}). In the semi--covariant
representation Eq.~(\ref{lorentztrick}) the origin of the phase factor $(-1)^{J+J'}$ is 
clearly identified. Also, Eq.~(\ref{lorentztrick}) does not depend on the phase conventions 
chosen for the polarization four-vectors since they always enter in squared form. 
This is different in the master formula in Eq.~(\ref{lepton-hadron2}) and the master formulas
written down in the following sections. They depend on
the correct choice of phases for the polarization four-vectors and for the matrix 
elements of Wigner's $d^J$--functions. Judging from the fact that there exist different 
conventions for these phases in the literature the reader can appreciate what a 
hazardous enterprise it can be to get all the signs correct in the angular decay 
distributions if one has to rely solely on master formulas without explication of 
phase conventions. Whereas the signs of the polar correlations can usually be checked 
by angular momentum considerations there is no easy way to check on the signs of the 
azimuthal correlations to be discussed in the subsequent sections. In fact, we have 
repeatedly used the semi--covariant representation Eq.~(\ref{lorentztrick}) to check on
the correctness of the phase conventions for the polarization four-vectors and Wigner's 
$d^J$--functions used in the different master formulas in this paper.

\section{Some numerical results} \label{section-numerics}
\subsection{Total semileptonic rates}

In order to obtain the total semileptonic rate we numerically integrate the differential
rate Eq.~(\ref{diffrateq^2}) over $q^{2}$ in the range 
$m_{l}^{2} \leqslant q^{2} \leqslant (M_{1}-M_{2})^{2}$
using the minimal model form factors described in Sec.~\ref{section-helicity-ampt}. For the
$e$-mode one obtains
\begin{equation}
\label{ratee}
\Gamma(\Xi^0 \to \Sigma^+ + e^- + \bar{\nu}_e)=|V_{us}|^2\,\, 1.169 \cdot 10^{-17}\, {\rm GeV}
\end{equation}
This translates into a branching ratio of
\begin{equation}
\label{bre}
BR(\Xi^0 \to \Sigma^+ + e^- + \bar{\nu}_e)= 2.513 \cdot 10^{-4} 
\end{equation}
where we have used central PDG values for the lifetime of the $\Xi^0$ 
($\tau(\Xi^0)= 2.90 \cdot 10^{-10}s$) 
and the CKM matrix element $V_{us}$ ($V_{us}=0.2255$) \cite{pdg08}.
The branching ratio (\ref{bre}) agrees very well with the experimental branching ratio
$BR(\Xi^0 \to \Sigma^+ + e^- + \bar{\nu}_e)= (2.53\pm0.08)\cdot 10^{-4}$ \cite{pdg08}.

For the $\mu$--mode we obtain
\begin{equation}
\label{ratemu}
\Gamma(\Xi^0 \to \Sigma^+ + \mu^- + \bar{\nu}_{\mu})
=|V_{us}|^2\,\,9.893 \cdot 10^{-20}\,{\rm GeV}\,,  
\end{equation}
which corresponds to a branching ratio of 
\begin{equation}
\label{brmu}
BR(\Xi^0 \to \Sigma^+ + \mu^- + \bar{\nu}_{\mu})=2.127 \cdot 10^{-6}\,.
\end{equation}
In the PDG listings one finds 
$BR(\Xi^0 \to \Sigma^+ + \mu^- + \bar{\nu}_{\mu})=(4.6 \,{+1.8 \atop -1.4})\, 
\cdot 10^{-6}$ \cite{pdg08}.
The NA48 collaboration has published
a preliminary result on this branching ratio which reads
$(2.2\pm0.3\pm0.2)\cdot 10^{-6}$ (102 events) \cite{lazzeroni05,wanke07} using a much larger
data sample than that of the KTeV Collaboration (nine events) \cite{ktev05} on which the 
PDG value is based on. Our branching ratio (\ref{brmu}) nicely agrees with the NA48
value but disagrees with the KTeV value by more than one standard deviation.

As it turns out one can get quite close to the exact results on the total semileptonic
rates in a simplified setting. The resulting formulas are quite useful for a first
assessment of the dominant dynamics of semileptonic hyperon decays. First, one 
neglects the contributions of 
the form factors $F_{2,3}^{V,A}$ since they contribute at most at 
$O((M_{1}-M_{2})/M_{1})$.
For the remaining two form factors $F_1^{V}$ and $F_1^{A}$ we neglect the
$q^{2}$--dependence since their $q^{2}$--variation in the allowed $q^{2}$--range is
is only of $O(2\%)$ (see Sec.~\ref{section-helicity-ampt}). With these simplifying 
assumptions the $q^{2}$--integration of the differential rate in Eq.~(\ref{diffrateq^2}) 
can be done analytically.
The result is again written in terms of the scaled variables  
$\rho^2=M_2^2/M_1^2$ and $\eta^2=m_l^2/M_1^2$, and the form 
$L=(1+\rho^4+\eta^4-2\rho^2-2\eta^2-2\rho^2 \eta^2)^{1/2}$. One has
\begin{equation}
\label{notequal}
\Gamma/\Gamma_0= \frac{1}{2} \bigg( |F_1^V|^2 R(\rho,\eta) +
|F_1^A|^2 R(-\rho,\eta) \bigg) \, , 
\end{equation}
where 
\begin{eqnarray}
\label{bigr}
R(\rho,\eta) &=& L\Big(A+A(\rho \leftrightarrow \eta) - B\Big)\nonumber\\
               && -24 \rho^3 \Big((1-\eta^2)^2 + \rho^2 + \rho(1-\eta^4)\Big) \ln E \nonumber \\
              && -24 \eta^4 (1- \rho^2)(1-\rho+\rho^2)
\ln E(\rho \leftrightarrow \eta) \, ,
\end{eqnarray}
and where we have used the abbreviations
\begin{eqnarray}
A &=& \frac{1}{2}-7\rho^2-7\rho^4+\rho^6+6 \eta^2\rho^2-7\eta^2\rho^4\,,  \nonumber \\
B &=& 2\rho (1 - 5 \eta^2 -2\eta^4 + 10\rho^2 - 5\rho^2 \eta^2 + \rho^4) \,,\nonumber \\
E &=& \frac{1+\rho^2-\eta^2 - L}{2|\rho|}\,. 
\end{eqnarray}
Upon setting $F_1^V=F_1^A=1$ in Eq.~(\ref{notequal}) one reproduces the SM
rate in Eq.~(\ref{smrate}).

Numerically, one has $BR(\Xi^0 \to \Sigma^+ + e^- + \bar{\nu}_e)=2.451\cdot 10^{-4}$ and
$BR(\Xi^0 \to \Sigma^+ + \mu^- + \bar{\nu}_{\mu})=2.061\cdot 10^{-6}$ using the 
approximate result Eq.~(\ref{notequal}) which is off the exact result in  
Eqs.~(\ref{bre}) and (\ref{brmu}) by $-2.4\%$ and $-3.1\%$, respectively. These
deviations from the exact model figures provide a measure of the importance of the form 
factors $F_{2}^{V}$ and $F_{3}^{A}$, and the $q^{2}$--dependence of the form factors.

Considering the fact that $(M_{1}-M_{2})$ is small compared to $(M_{1}+M_{2})$ we
expand Eq.~(\ref{notequal}) in the variable 
$\delta=(M_{1}-M_{2})/(M_{1}+M_{2})=(1-\rho)/(1+\rho)$ where
$\delta=0.0501$ in the case discussed in this paper. One obtains (see also \cite{p83})
\begin{eqnarray}
\label{approx}
\Gamma/\Gamma_0&=& \frac{512}{5}\frac{1}{(1+\delta)^{8}} \left[ 
|F_1^V|^2\,\Big(1+3\,\frac{|F_1^A|^{2}}{|F_1^V|^{2}}\Big)\,r(x)\,\,\delta^{5} + 
O(\delta^7)\right] \nonumber \\ 
&=&\frac{2}{5} (1-\rho)^{5}(1+\rho)^{3}
|F_1^V|^2\left(1+3\,\frac{|F_1^A|^{2}}{|F_1^V|^{2}}\right)r(x) + O\big((1-\rho)^7\big) \, .
\end{eqnarray} 

The coefficient of the remaining $\delta^{7}$--term and the higher order terms in the 
small $\delta$ expansion Eq.~(\ref{approx}) do not have as simple a structure as the 
remaining terms in Eq.~(\ref{approx}) but can be calculated to be quite small
($0.1\%$ and $0.3\%$, respectively, in the $e$-- and $\mu$--mode in the present case). It is
curious that the next-to-next correction to
 Eq.~(\ref{approx}) is 
$O(\delta^{9})$.  
The function $r(x)$ in Eq.~(\ref{approx}) is given by 
\begin{equation}
\label{p2}
r(x)=\frac{\sqrt{1-x^2}}{2}\left(2-9x^2-8x^4\right)- \frac{15}{2} x^4
\ln \frac{1-\sqrt{1-x^2}}{x} \, \, ,
\end{equation}
where $x=\eta/(1-\rho)$. The overall factor $(1+\delta)^{-8}$ in Eq.~(\ref{approx})
can be seen to arise from power counting in Eq.~(\ref{bigr}) considering the fact
that $\rho$ and $\eta$ are proportional to $(1+\delta)^{-1}$.   

The small $\delta$--expansion Eq.~(\ref{approx}) is quite
remarkable on two accounts. First, it is quite accurate since the corrections to 
Eq.~(\ref{approx}) set in only
at $O(\delta^{7})$ and not at $O(\delta^{6})$ as one would naively expect, i.e.
the corrections to Eq.~(\ref{approx}) are only of $O(\delta^{2}=0.00251)$. 
From Eq.~(\ref{approx}) one can quickly appreciate that the rate measurement is quite 
sensitive to the ratio $F_{1}^{A}/F_{1}^{V}$.
Second, it is quite remarkable that
Eq.~(\ref{approx}) factorizes into a lepton mass independent and
lepton mass dependent term. This will be very useful for a quick estimate of
the rate ratio $\Gamma(e)/\Gamma(\mu)$ to be discussed in the next subsection.

\subsection{The rate ratio $\Gamma(e)/\Gamma(\mu)$}
Of interest is the rate ratio $\Gamma(e)/\Gamma(\mu)$ which has been measured
by the KTeV collaboration. Dividing the $e$-mode rate in Eq.~(\ref{ratee}) by 
the $\mu$--mode rate in Eq.~(\ref{ratemu}) we obtain 
\begin{equation}
\label{rateratio1}
\Gamma(e)/\Gamma(\mu)= 118.13\;\;(55.6^{+22.2}_{-16.7}\,\cite{ktev05} )\,.
\end{equation}
We have added the corresponding published experimental rate ratio and its 
errors from the KTeV collaboration in brackets. The KTeV value is off by more than 
two standard deviations from the calculated value in Eq.~(\ref{rateratio1}). We mention 
that the NA48 Collaboration cites a preliminary value of $114.1 \pm 19.4$  
for the rate ratio $\Gamma(e)/\Gamma(\mu)$ \cite{lazzeroni05,wanke07} which is in very good
agreement with the calculated result in Eq.~(\ref{rateratio1}).

Using the simplified form in Eq.~(\ref{notequal}) we calculate 
$\Gamma(e)/\Gamma(\mu)=118.96$ which is quite close to the full model value in
Eq.~(\ref{rateratio1}). 

Finally, we consider the small $\delta$--approximation of
Eq.~(\ref{notequal}) given by  Eq.~(\ref{approx}). Since the small $\delta$--approximation
factors into a lepton mass independent part and a lepton mass dependent part $r(x)$
the rate ratio $\Gamma(e)/\Gamma(\mu)$ is simply given by the ratio 
$r(x_{e})/r(x_{\mu})$.
In particular this shows that, at this level of approximation, the rate ratio does not
depend on the actual values of $F_1^V(0)$ and $F_1^A(0)$. This in turn implies that a 
measurement of the rate ratio $\Gamma(e)/\Gamma(\mu)$ does not reveal much of the 
dynamics of semileptonic hyperon decays except for a test of $e$--$\mu$--universality. 
Numerically, one obtains 
\begin{equation}
\label{rateratio2}
\Gamma(e)/\Gamma(\mu)= r(x_{e})/r(x_{\mu})= 118.71\,,
\end{equation}
which is quite close to the corresponding ratio of $118.96$ calculated in the 
simplified setting according to formula Eq.~(\ref{notequal}) as one
would expect from the quality of the small $\delta$--approximation Eq.~(\ref{approx}).



\subsection{Forward-backward asymmetry}
A very useful measure is the forward-backward asymmetry of the lepton in the
$W^-_{\rm{off-shell}}$ rest frame (or $(l,\nu_{l})$ c.m. frame) defined by
\begin{align}
\label{fb}
A_{FB}(q^{2}) & = \frac{d\Gamma/dq^{2}({\it forward})- d\Gamma/dq^{2}({\it backward})}
{d\Gamma/dq^{2}({\it forward})+ d\Gamma/dq^{2}({\it backward})}\nonumber\\
& := \frac{N(q^{2})}{D(q^{2})}\,.
\end{align}
The numerator factor can be calculated from  Eqs.~(\ref{diffrate3})
and (\ref{lepton-hadron1}) and reads
\begin{align}
N(q^{2}) =& \frac{G_F^2}{(2\pi)^3}
|V_{us}|^2 \frac{(q^2-m_l^2)^2 p}{8M_1^2q^2}\left[\mp H_{\frac{1}{2}1}^V
H_{\frac{1}{2}1}^A - 2\frac{m_{l}^{2}}{2q^{2}}(H_{\frac{1}{2}t}^V H_{\frac{1}{2}0}^V
+H_{\frac{1}{2}t}^A H_{\frac{1}{2}0}^A)\right]\,.\nonumber
\end{align}
The denominator factor is simply given by the differential rate Eq.~(\ref{diffrateq^2}),
i.e. $D(q^{2})= d\Gamma/dq^{2}$.

For $m_{l}=0$ (which is realized for the $e$--mode for all practical purposes) the 
forward-backward asymmetry can be seen to be directly proportional to the helicity 
amplitude $H_{\frac{1}{2}1}^V$ and is thus very sensitive to the ratio of form
factors $F_{2}^{V}/F_{1}^{V}$ as discussed in Sec.~\ref{section-helicity-ampt}. In the
$\mu$--mode there is in addition some sensitivity to the helicity amplitude 
$H_{\frac{1}{2}t}^A$ which implies a certain sensitivity to the form factor ratio
$F_{3}^{A}/F_{1}^{A}$. Looking at the size and signs of the helicity amplitudes in
Fig.~\ref{helamp} one finds that the forward-backward asymmetry is positive for the
$\Xi^0 \to \Sigma^+$ transition in the $e$--mode and negative for the $\mu$--mode. 

When averaging $A_{FB}(q^{2})$ over $q^{2}$ the $q^{2}$ integration has to be done 
separately in the numerator and the denominator of Eq.~(\ref{fb}). Using again the
minimal form factor model of Sec.~\ref{section-helicity-ampt} we obtain
\begin{equation}
\label{afbe}
\langle A_{FB} \rangle(e{\rm -mode}) = 0.081 \,\, (0.014)\,,
\end{equation}
where we have added the corresponding value with $F_{2}^{V}$ switched off in brackets. 
The big difference between the two numbers in Eq.~(\ref{afbe}) underscores the sensitivity 
of the forward--backward measure to the form factor ratio $F_{2}^{V}/F_{1}^{V}$.
For the $\mu$--mode one obtains
\begin{equation}
\label{afbmu}
\langle A_{FB} \rangle(\mu{\rm -mode}) = -0.082 \,\, (-0.115;-0.087)\,, 
\end{equation}
where the two figures in brackets refer to $F_{2}^{V}$ and $F_{3}^{A}$, respectively, being
switched off. There still is a sensitivity to the ratio $F_{2}^{V}/F_{1}^{V}$, but
at a much reduced level compared to the $e$--mode. The second number in the brackets
of Eq.~(\ref{afbmu}) indicates that there is a very small sensitivity to the ratio 
$F_{3}^{A}/F_{1}^{A}$.
\section{Single spin polarization effects} \label{section-single-spin}

\subsection{Polarization of the daughter baryon}
The lepton-hadron contraction $L_{\mu\nu}H^{\mu\nu}$ given in Eqs.~(\ref{lepton-hadron1})
and (\ref{lepton-hadron2}) 
can be separated into
contributions of positive and negative helicities of the daughter baryon 
denoted by $L_{\mu\nu}H^{\mu\nu}_{\pm\pm}$. They are given by
\begin{align}
L_{\mu\nu}H^{\mu\nu}_{++}(\theta)=&\frac{2}{3}(q^2-m_l^2) \;
\bigg[  \frac{3}{8}(1\mp\cos \theta)^2
|H_{\frac{1}{2}1}|^2 +\; \frac{3}{4} \sin^2 \theta |H_{\frac{1}{2}0}|^2  \nonumber \\
& +\; \frac{m_l^2}{2q^2} \;
\bigg\{
\frac{3}{2}|H_{\frac{1}{2}t}|^2 
+\frac{3}{4}|H_{\frac{1}{2}1}|^2 \sin^2\theta \nonumber \\
& +\; \frac{3}{2}\cos^2 \theta|H_{\frac{1}{2}0}|^2 
-3\cos \theta H_{\frac{1}{2}t}H_{\frac{1}{2}0}  \bigg\} \;\bigg]\,,\\[2mm]
L_{\mu\nu}H^{\mu\nu}_{--}(\theta)=&\frac{2}{3}(q^2-m_l^2) \;
\bigg[ 
 \frac{3}{8}(1\pm\cos \theta)^2 |H_{-\frac{1}{2}-1}|^2 
+\; \frac{3}{4} \sin^2 \theta  |H_{-\frac{1}{2}0}|^2 \nonumber \\
& + \frac{m_l^2}{2q^2} \;
\bigg\{
\frac{3}{2} |H_{-\frac{1}{2}t}|^2
+\frac{3}{4} |H_{-\frac{1}{2}-1}|^2 \sin^2\theta  \nonumber \\
& +\; \frac{3}{2}\cos^2 \theta |H_{-\frac{1}{2}0}|^2
-3\cos \theta  
H_{-\frac{1}{2}t}H_{-\frac{1}{2}0} \bigg\} \;\bigg]\,.
\end{align}

This allows one to compute the component $P_z$ of the polarization vector 
along the direction of $\vec{p}_2$ in the rest system of $B_1$. One obtains
\begin{equation}
\label{polz}
P_z(\theta) = \frac{L_{\mu\nu}H^{\mu\nu}_{++} - L_{\mu\nu}H^{\mu\nu}_{--}}
{L_{\mu\nu}H^{\mu\nu}_{++} + L_{\mu\nu}H^{\mu\nu}_{--}}\,.
\end{equation}
On average\footnote{When averaging  $P_z$
over $q^{2}$ one has to separately integrate the numerator and denominator of 
Eq.~(\ref{polz}) after restoring the factor $p(q^{2}-m_{l}^{2})/q^{2}$ 
in both the numerator and the denominator.} 
one has $\langle P_{z}\rangle =-0.65$ and $\langle P_{z}\rangle =-0.33$, 
respectively, for the $e^{-}$--mode and $\mu^{-}$--modes.

In a similar vein the polarization of the daughter baryon in the $x$--direction can be 
obtained from Eq.~(\ref{lepton-hadron2}) by leaving the helicity label $\lambda_2$  
unsummed. The product of helicity amplitudes now reads 
$H_{\lambda_2 \lambda_W}H_{\lambda'_2 \lambda'_W}^\ast$ and the Kronecker $\delta$ turns
into $\delta_{\lambda_2-\lambda_W,\,\lambda'_2-\lambda'_W}$ because, again,
the parent baryon is taken to be unpolarized. As before, $\lambda_W=t$ and $\lambda_W=0$
have zero helicity but transform as $J=1$ and $J=0$, respectively. One obtains 
\begin{equation}
\label{polx}
P_x(\theta) = \frac{2 L_{\mu\nu}H^{\mu\nu}_{+-}}
{L_{\mu\nu}H^{\mu\nu}_{++} + L_{\mu\nu}H^{\mu\nu}_{--}},
\end{equation}
where
\begin{align}
\label{transverse-p}
2 L_{\mu\nu} H^{\mu\nu}_{+-}(\theta)=& -\frac{2}{3}(q^2-m_l^2) 
\bigg[\frac{3}{2\sqrt{2}}\sin\theta(\pm1-\cos\theta)H_{\frac{1}{2}1}H_{-\frac{1}{2}0}\nonumber\\ 
& + \frac{3}{2\sqrt{2}}\sin\theta(\pm1+\cos\theta)H_{\frac{1}{2}0}
H_{-\frac{1}{2}-1}  \nonumber \\ 
&+\frac{m_l^2}{2q^2}\Big\{\frac{3}{\sqrt{2}}\sin\theta \cos\theta
(H_{\frac{1}{2}1}H_{-\frac{1}{2}0} -H_{\frac{1}{2}0}H_{-\frac{1}{2}-1})\nonumber\\ 
& -\frac{3}{\sqrt{2}}\sin\theta
(H_{\frac{1}{2}1}H_{-\frac{1}{2}t} - H_{\frac{1}{2}t}H_{-\frac{1}{2}-1})
\Big\} \bigg] \;.
\end{align}
Of course, if one does not define a transverse reference direction the specification 
of $P_x$ does not make physical sense per se. Such a transverse reference direction is
e.g. provided by the transverse momentum of the lepton in the semileptonic
decay. In fact, we shall see in Sec.~\ref{section-joint} how the density matrix of the daughter
baryon enters the joint angular decay distribution of the cascade decay
$\Xi^0 \to \Sigma^+ (\to p + \pi^+) + l^- + \bar{\nu}_l$ where the transverse
reference direction is defined by the decay $\Sigma^+ \to p + \pi^+$.    
The polarization component $P_y$ is zero because we assume that the invariant
amplitudes and thereby the helicity amplitudes are relatively real. 
On average one has $\langle P_{x}\rangle =-0.57$ and $\langle P_{x}\rangle =-0.17$, 
respectively, for the $e^{-}$--mode and $\mu^{-}$--modes.

\subsection{Polarization of the lepton}
The lepton--side flip-- and nonflip--contributions to $L_{\mu\nu}H^{\mu\nu}$ are 
clearly identifiable as can be seen by an inspection of Eqs.~(\ref{lepton-hadron2}) and
(\ref{diffrateq^2}). One can thus directly write down the longitudinal
polarization of the lepton for the decay of an unpolarized parent hyperon at no extra 
cost. One has

\begin{equation}
\label{leplongpol}
P_z^{(l)} = \pm \frac{L_{\mu\nu}H^{\mu\nu}({\it flip})-L_{\mu\nu}H^{\mu\nu}({\it nonflip})}
{L_{\mu\nu}H^{\mu\nu}({\it flip})+L_{\mu\nu}H^{\mu\nu}({\it nonflip})}\,.
\end{equation} 

\begin{figure}[ht!]
  \begin{center}
    \epsfig{figure=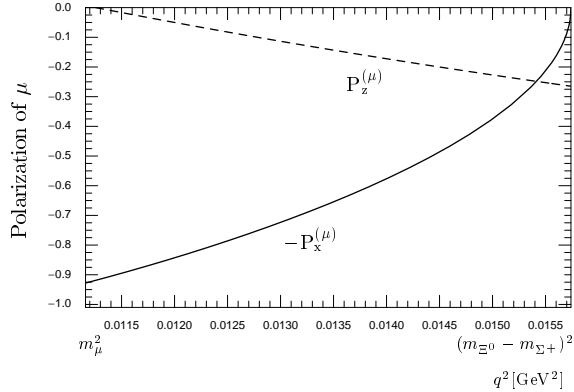, width=7.5cm}
     \caption{\label{polmu} \small \sf
 Longitudinal and transverse polarization of the $\mu^-$
in the $(\mu^-,\bar{\nu}_\mu)$ c.m. system.}
  \end{center}
\end{figure}

For the decay $\Xi^0 \to \Sigma^+ + l^- + \bar{\nu}_l$ the longitudinal polarization of the 
electron is $\approx- 100\%$ over most of the range of $q^2$ because $m_e \approx 0$. 
This changes only for $q^2$--values very close to the threshold $q^2=m_e^2$. For the 
$\mu^-$--mode the longitudinal polarization is quite small and negative and remains smaller 
than $\sim 30\%$ in magnitude  over the whole $q^2$--range as Fig.~\ref{polmu} shows. 
On average one has
$\langle P^{(\mu^-)}_z \rangle=-0.18$. Judging from the fact that $P_z^{(\mu^-)}$ is small 
the helicity flip and nonflip contributions are of almost equal importance for the 
$\mu^-$--mode.

It is important to realize that the longitudinality of the polarization 
$P_z^{(l)}$ is defined with respect to the momentum direction of the lepton in the
$(l,\nu_l)$ c.m. system and {\it not} with respect to the momentum direction of the 
lepton in the rest system of the parent baryon $\Xi^0$. If one needs to
avail of the longitudinal polarization in the latter frame this can also be done using
the helicity method as has been shown in \cite{hmw90}.

As before, the transverse polarization of the lepton can also be obtained 
from Eq.~(\ref{lepton-hadron2}) by leaving the helicity label $\lambda_l$
in Eq.~(\ref{lepton-hadron2}) unsummed. One then obtains the density matrix of the lepton
which we write as $(L_{\mu\nu}H^{\mu\nu})_{\lambda_l \lambda_{l'}}$. This allows one to 
extract also the transverse polarization of the lepton $P_x^{(l)}$.
One obtains (see also \cite{bkp88})
\begin{equation}
\label{polxmu}
P_x^{(l)}(\theta) = \frac{2 (L_{\mu\nu}H^{\mu\nu})_{+-}}
{(L_{\mu\nu}H^{\mu\nu})_{++} + (L_{\mu\nu}H^{\mu\nu})_{--}}\,.
\end{equation}

In order to evaluate Eq.~(\ref{polxmu}) for the $(l^-,\bar{\nu})$ case one needs the relation 
$h_{\frac{1}{2}\frac{1}{2}}=\sqrt{m_l^2/2q^2}\,\,h_{-\frac{1}{2}\frac{1}{2}}$.
In Fig.~\ref{polmu} we show the $q^2$--dependence of the transverse polarization of 
the $\mu^-$ in the decay $\Xi^0 \to \Sigma^+ + l^- + \bar{\nu}_l$. The transverse 
polarization starts off at rather high
positive values close to $q^2_{\rm{min}}=m_\mu^2$ and drops to zero at the zero
recoil point $q^2_{\rm{max}}=(M_1-M_2)^2$. For the $e^-$ the transverse polarization
is practically zero over the whole $q^2$--range. Because of the lack of structure in the
$e^-$--case we do not show a plot of the polarization of the electron. 
\subsection{Decay of a polarized parent baryon}
In this subsection we consider the decay of a polarized parent baryon and in turn 
determine 
the angular decay distributions of the leptonic side and the hadronic side relative
to the polarization of the parent baryon. The polarization of the parent
baryon is described by the density matrix
\begin{equation}
\label{densitym}
\mathrm{\rho_{\lambda_{1}\lambda_{1}^{'}}(\theta_{P})}=\frac{1}{2}\left(
  \begin{array}{cc}
    1+\mathrm{P} \cos{\theta_{\mathrm{P}}} & \mathrm{P} \sin{\theta_{\mathrm{P}}}\\
    \mathrm{P} \sin{\theta_{\mathrm{P}}} & 1-\mathrm{P} \cos{\theta_{{\mathrm{P}}}}
   \end{array}
\right)\,,
\end{equation}
where we have assumed that the polarization vector of the parent baryon lies in the 
$(x,z)$--plane with positive $x$--component as shown in Figs.~\ref{lepton-side} and 
\ref{hadron-side}. The rows and columns of the matrix in Eq.~(\ref{densitym}) 
are labeled in the order $(1/2,-1/2)$.

\subsubsection{Lepton side as polarization analyzer}

\begin{figure}[t]\begin{center}
\epsfig{figure=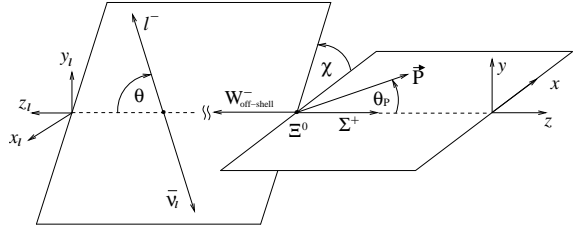, width=7.5cm}
\caption{\label{lepton-side} \small \sf
Definition of the polar angles $\theta$ and $\theta_P$,
and the azimuthal angle $\chi$ describing the decay of a polarized $\Xi^0$ 
using the lepton side as polarization analyzer. 
$\vec{P}$ denotes the polarization vector of the $\Xi^0$.
The coordinate system $(x_l, y_l, z_l)$ is obtained from the 
coordinate system $(x,y,z)$ by a $180^\circ$ rotation around the $y$--axis. }
\end{center}\end{figure}

The angular decay distribution is a straightforward generalization of 
Eq.~(\ref{lepton-hadron1}) where one now has to include the density matrix of the
decaying parent baryon $B_1$. Also, the rotation of the density matrix of the
$W_{\rm{off-shell}}$ into the direction of the lepton now involves also the 
azimuthal angle $\chi$. This brings in the phase factor 
$e^{i(\lambda_W-\lambda'_W)(\pi-\chi)}$. The appropriate angle entering the phase factor
is $(\pi-\chi)$ since the azimuthal angle has to be
specified in the leptonic $(x_l,y_l)$--plane (see Fig.~\ref{lepton-side}). 
Using Appendix~\ref{appendix-2body-master}, one obtains the master formula
\begin{align}
\label{lside1}
W(\theta,\chi,\theta_P)  \propto& \sum_{\lambda_l,\lambda_2,\lambda_W,\lambda'_W,J,J'} 
\rho_{\lambda_2-\lambda_W,\lambda_2-\lambda'_W}(\theta_P) (-1)^{J+J'}
|h^l_{\lambda_l\lambda_{\nu}=\pm1/2}|^2
e^{i(\lambda_W-\lambda'_W)(\pi-\chi)} \times \nonumber \\ 
&\times d^J_{\lambda_W,\lambda_l-\lambda_{\nu}}(\theta)
d^{J'}_{\lambda'_W,\lambda_l-\lambda_{\nu}}(\theta)
H_{\lambda_2\lambda_W}H^*_{\lambda_2\lambda'_W} \,,
\end{align}
where $\lambda_{\nu}=\pm1/2$ ($\lambda_{\nu}=1/2$ for $(l^-,\bar{\nu}_l)$ and
$\lambda_{\nu}=-1/2$ for $(l^+,\nu_l)$).

Doing the helicity sums and putting in the correct normalization one obtains 
{
\allowdisplaybreaks
\begin{align}
\label{lside2}
& \frac{d\Gamma}{dq^2 d\cos\theta d\chi d\cos\theta_P}=
\frac{1}{6}\frac{G_F^2}{(2\pi)^4}|V_{us}|^2
\frac{(q^2-m_l^2)^2p}{8M_1^2q^2}\times \nonumber \\ \nonumber
& \quad \times \bigg[\,  \frac{3}{8}(1\mp\cos\theta)^2 |H_{\frac{1}{2}1}|^2 
(1-P\cos\theta_P)  +\frac{3}{8}(1\pm\cos\theta)^2 |H_{-\frac{1}{2}-1}|^2 
(1+P\cos\theta_P) \\ \nonumber
&\quad  +\frac{3}{4}\sin^2\theta\Big(
|H_{\frac{1}{2}0}|^2(1+P\cos\theta_P) +|H_{-\frac{1}{2}0}|^2(1-P\cos\theta_P)\Big) \\ \nonumber
&\quad \pm\frac{3}{2\sqrt{2}}P \sin\theta\cos\chi \sin\theta_P
\Big((1\mp\cos\theta)H_{\frac{1}{2}1}H_{\frac{1}{2}0}  
+(1\pm\cos\theta)H_{-\frac{1}{2}-1}H_{-\frac{1}{2}0}\Big) \\
\nonumber
&\quad +\frac{m_l^2}{2q^2} \bigg\{ \, \frac{3}{2}|H_{\frac{1}{2}t}|^2
(1+P\cos\theta_P)
+\frac{3}{2}|H_{-\frac{1}{2}t}|^2(1-P\cos\theta_P) \\ \nonumber
&\quad -3\cos\theta \Big(H_{\frac{1}{2}t}H_{\frac{1}{2}0}
(1+P\cos\theta_P)
+H_{-\frac{1}{2}t}H_{-\frac{1}{2}0}(1-P\cos\theta_P)\Big) \\ \nonumber
&\quad +\frac{3}{2}\cos^2\theta \Big(|H_{\frac{1}{2}0}|^2
(1+P\cos\theta_P)
+|H_{-\frac{1}{2}0}|^2(1-P\cos\theta_P)\Big) \\ \nonumber
&\quad +\frac{3}{4}\sin^2\theta\Big(|H_{\frac{1}{2}1}|^2
(1-P\cos\theta_P)
+|H_{-\frac{1}{2}-1}|^2(1+P\cos\theta_P)\Big) \\ \nonumber
&\quad -\frac{3}{\sqrt{2}}P \sin\theta \cos\chi \sin\theta_P 
(H_{\frac{1}{2}1}H_{\frac{1}{2}t} - H_{-\frac{1}{2}-1}H_{-\frac{1}{2}t})
\\
&\quad +\frac{3}{\sqrt{2}}P
\sin\theta \cos\theta \cos\chi \sin\theta_P  (H_{\frac{1}{2}1}H_{\frac{1}{2}0} - H_{-\frac{1}{2}-1}H_{-\frac{1}{2}0})
\bigg\}\bigg] \,.
\end{align}
}
A similar result was published in \cite{ft71}. However, 
our result in Eq.~(\ref{lside2}) does not agree with the corresponding result
in \cite{ft71}.

\subsubsection{Hadron side as polarization analyzer}
Following the familiar procedure of building up the cascade decay in a quasi-factorized
form one obtains the master formula 

\begin{align}
\label{hside1}
W (\theta_B,\phi_B,\theta_P) \propto& \sum_{\lambda_l,\lambda_W,\lambda'_W,J,J',\lambda_2,
\lambda_2',\lambda_3}
(-1)^{J+J'} 
\rho_{\lambda_2-\lambda_W,\lambda'_2-\lambda'_W}(\theta_P)
H_{\lambda_2\lambda_W}H^*_{\lambda'_2\lambda'_W}\times  \nonumber \\
&\times \int^{2\pi}_0 d\phi_l \int^1_{-1} d\cos\theta
\,\,|h^l_{\lambda_l\lambda_{\nu}=\pm1/2}|^2 e^{i(\lambda_W-\lambda'_W)\phi_l}\times \nonumber\\
&\times  d^J_{\lambda_W,\lambda_l-\lambda_{\nu}}(\theta)d^{J'}_{\lambda'_W,\lambda_l-\lambda_{\nu}}
(\theta) e^{i(\lambda_2-\lambda'_2)\phi_B} d^{\frac{1}{2}}_{\lambda_2\lambda_3}(\theta_B)
d^{\frac{1}{2}}_{\lambda_2'\lambda_3}(\theta_B)
\,\,|h^B_{\lambda_3 0}|^2 \,\, ,
\end{align}
where the $h^B_{\lambda_3 0}$ are the helicity amplitudes of the decay 
$B_2 \to B_3 + \pi$. Latter decay is as usual characterized by the asymmetry parameter
\begin{equation}
\alpha_B = \frac{|h^B_{\frac{1}{2} 0}|^2 - |h^B_{-\frac{1}{2} 0}|^2}
{|h^B_{\frac{1}{2} 0}|^2 + |h^B_{-\frac{1}{2} 0}|^2} \,\, .
\end{equation}
\begin{figure}[t]\begin{center}
\epsfig{figure=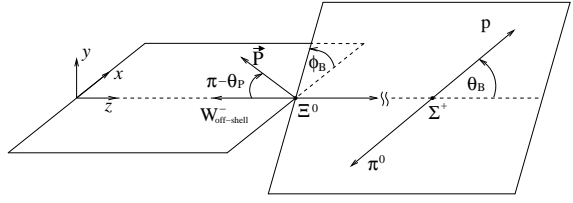, width=7.5cm}
\caption{\label{hadron-side}\small \sf
 Definition of the polar angles $\theta_B$ and $\theta_B$
and the azimuthal angle $\phi_B$ in the cascade decay of a polarized $\Xi^0$ 
using the hadron side as polarization analyzer. }
\end{center}
\end{figure}
The asymmetry parameter for the nonleptonic decay $\Sigma^+ \to p + \pi^0$ relevant
to this paper is given by $\alpha_B = -0.980{+0.017 \atop -0.015}$ \cite{pdg08}.
Note that the phase factor in Eq.~(\ref{hside1}) now is 
$\rm{exp}[i(\lambda_2-\lambda'_2)\phi_B]$ which is appropriate for the azimuthal angle
$\phi_B$ measured relative to the $(x,z)$--plane (see Fig.~\ref{hadron-side}).

Doing the helicity sum and the integration in Eq.~(\ref{hside1}), and putting in
the correct normalization one obtains
\begin{align}
\label{hside2}
\frac{d\Gamma}{dq^2 d\!\cos\theta_B d\phi_B d\!\cos\theta_P} =&
B(B_2 \to B_3 + \pi) \frac{1}{12}\frac{G_F^2}{(2\pi)^4}|V_{us}|^2
\frac{(q^2-m_l^2)^2p}{8M_1^2q^2}\times  \\ \nonumber
&\times \bigg[ (1+\frac{m_l^2}{2q^2})(1+\alpha_B\cos\theta_B)
(1-P \cos\theta_P) |H_{\frac{1}{2}1}|^2 \\ \nonumber 
& +(1+\frac{m_l^2}{2q^2})(1-\alpha_B\cos\theta_B)
(1+P \cos\theta_P) |H_{-\frac{1}{2}-1}|^2 \\ \nonumber
& + 
(1+\frac{m_l^2}{2q^2})(1+\alpha_B\cos\theta_B)
(1+P \cos\theta_P) |H_{\frac{1}{2}0}|^2 \\ \nonumber
&+ (1+\frac{m_l^2}{2q^2})(1-\alpha_B\cos\theta_B)
(1-P \cos\theta_P) |H_{-\frac{1}{2}0}|^2 \\ \nonumber
&+2P \alpha_B \sin\theta_B\cos\phi_B\sin\theta_P 
H_{\frac{1}{2}0}H_{-\frac{1}{2}0} \\ \nonumber
 &+\frac{m_l^2}{2q^2} \bigg\{
(1+\alpha_B\cos\theta_B)(1+P \cos\theta_P)\,3\,|H_{\frac{1}{2}t}|^2
\\ \nonumber
&+
(1-\alpha_B\cos\theta_B)(1-P \cos\theta_P)\,3\,|H_{-\frac{1}{2}t}|^2 \\ 
&+2P \alpha_B \sin\theta_B\cos\phi_B\sin\theta_P
(H_{\frac{1}{2}0}H_{-\frac{1}{2}0} +3 H_{\frac{1}{2}t}H_{-\frac{1}{2}t}) 
 \bigg\} \bigg] \,\, ,
\end{align}
where $B(B_2 \to B_3 + \pi)$ is the branching fraction of the nonleptonic decay 
$B_2 \to B_3 + \pi$.

\section{Joint angular decay distribution} \label{section-joint}

Following the familiar procedure the joint angular decay distribution for the
semileptonic cascade decay $B_1 \to B_2 (\to B_3 + \pi) + l + \nu_l$ of an unpolarized
parent baryon $B_1$ can be derived 
from the master formula\footnote{Much to
the embarrassment of one of the present authors (JGK) there was a sign mistake in the
azimuthal correlation term of the corresponding joint angular decay distribution written 
down in \cite{Korner:1991ph} for the
semileptonic decay $\Lambda_c \to \Lambda (\to p +  \pi^-)+l^+ + \nu_l$ $(m_l=0)$. The 
source of this error was that in \cite{Korner:1991ph} we used the phase
factor $\rm{exp}\,[i(\lambda_W-\lambda'_W)(-\chi)]$ instead of the correct form
$\rm{exp}\,[i(\lambda_W-\lambda'_W)(\pi-\chi)]$ to determine the sign of the azimuthal
correlation term. This error was discovered and rectified through  
experimental evidence \cite{Hinson:2004pj}.}
\begin{align}
\label{joint1}
W(\theta,\chi,\theta_B)  \propto & \sum_{\lambda_l,\lambda_W,\lambda'_W,J,J',\lambda_2,
\lambda_2',\lambda_3} (-1)^{J+J'}
|h^l_{\lambda_l\lambda_{\nu}=\pm1/2}|^2
e^{i(\lambda_W-\lambda'_W)(\pi-\chi)}\delta_{\lambda_2-\lambda_W,\lambda'_2-\lambda'_W}\times
 \nonumber \\ 
&
\times d^J_{\lambda_W,\lambda_l-\lambda_{\nu}}(\theta)d^{J'}_{\lambda'_W,\lambda_l-\lambda_{\nu}}
(\theta)
H_{\lambda_2\lambda_W}H^*_{\lambda'_2\lambda'_W}\,d^{\frac{1}{2}}_{\lambda_2\lambda_3}(\theta_B)
d^{\frac{1}{2}}_{\lambda_2'\lambda_3}(\theta_B)
|h^B_{\lambda_3 0}|^2 \,\,.
\end{align}
\begin{figure}[t]\begin{center}
\epsfig{figure=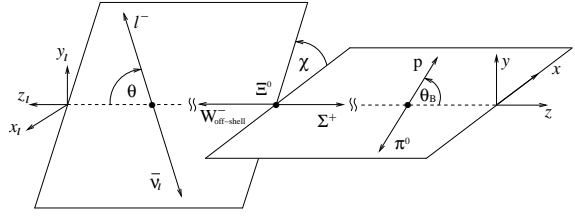, width=7.5cm}
\caption{\label{five-fold}\small \sf
Definition of the polar angles $\theta$ and $\theta_B$,
and the azimuthal angle $\chi$ in the joint angular decay distribution of an unpolarized
$\Xi^0$ in the cascade decay
$\Xi^0 \to \Sigma^+ (\to p + \pi^0) + l^- + \bar{\nu}_l$. The coordinate system
$(x_l, y_l, z_l)$ is obtained from the coordinate system $(x,y,z)$ by a $180^\circ$
rotation around the $y$--axis.}
\end{center}
\end{figure}
The polar angles $\theta$, $\theta_{B}$ and the azimuthal angle $\chi$ are defined in
Fig.~\ref{five-fold}.
The Kronecker $\delta$ in Eq.~(\ref{joint1}) expresses the fact that we are dealing with the
decay of an unpolarized parent hyperon which implies 
$\lambda_2-\lambda_W=\lambda'_2-\lambda'_W$.

When writing down the corresponding normalized decay distribution we shall as before
assume that the helicity amplitudes are relatively real. One obtains
{\allowdisplaybreaks 
\begin{align}
\label{joint2}
\frac{d\Gamma}{dq^2 d\cos\theta d\chi d\cos\theta_B} &=
B(B_2 \to B_3 + \pi)\frac{1}{6}\frac{G_F^2}{(2\pi)^4}|V_{us}|^2
\frac{(q^2-m_l^2)^2p}{8M_1^2q^2}\times\nonumber\\
& \hspace*{-2cm}\times\bigg[ \frac{3}{8}(1\mp\cos\theta)^2 |H_{\frac{1}{2}1}|^2 
(1+\alpha_B\cos\theta_B) +\frac{3}{8}(1\pm\cos\theta)^2 |H_{-\frac{1}{2}-1}|^2 
(1-\alpha_B\cos\theta_B) \nonumber\\
&\hspace*{-2cm}\quad +\frac{3}{4}\sin^2\theta\Big(
|H_{\frac{1}{2}0}|^2(1+\alpha_B\cos\theta_B)
+|H_{-\frac{1}{2}0}|^2(1-\alpha_B\cos\theta_B)\Big)  \nonumber\\ 
&\hspace*{-2cm}\quad \pm\frac{3}{2\sqrt{2}}\alpha_B \sin\theta \cos\chi \sin\theta_B
\Big((1\mp\cos\theta)H_{-\frac{1}{2}0}H_{\frac{1}{2}1}  
+(1\pm\cos\theta)H_{\frac{1}{2}0}H_{-\frac{1}{2}-1}\Big)\nonumber \\
&\hspace*{-2cm}\quad +\frac{m_l^2}{2q^2} \bigg\{\frac{3}{2}|H_{\frac{1}{2}t}|^2
(1+\alpha_B\cos\theta_B)
+\frac{3}{2}|H_{-\frac{1}{2}t}|^2(1-\alpha_B\cos\theta_B)\nonumber \\ 
&\hspace*{-2cm}\quad -3\cos\theta \Big(H_{\frac{1}{2}t}H_{\frac{1}{2}0}
(1+\alpha_B\cos\theta_B)
+H_{-\frac{1}{2}t}H_{-\frac{1}{2}0}(1-\alpha_B\cos\theta_B)\Big)\nonumber \\ 
&\hspace*{-2cm}\quad +\frac{3}{2}\cos^2\theta \Big(|H_{\frac{1}{2}0}|^2
(1+\alpha_B\cos\theta_B)
+|H_{-\frac{1}{2}0}|^2(1-\alpha_B\cos\theta_B)\Big) \\ \nonumber
&\hspace*{-2cm}\quad+\frac{3}{4}\sin^2\theta\Big(|H_{\frac{1}{2}1}|^2
(1+\alpha_B\cos\theta_B)
+|H_{-\frac{1}{2}-1}|^2(1-\alpha_B\cos\theta_B)\Big) \\ \nonumber
&\hspace*{-2cm}\quad-\frac{3}{\sqrt{2}}\alpha_B \sin\theta \cos\chi \sin\theta_B 
(H_{-\frac{1}{2}t}H_{\frac{1}{2}1} - H_{\frac{1}{2}t}H_{-\frac{1}{2}-1}) \\
&\hspace*{-2cm}\quad +\frac{3}{\sqrt{2}}\alpha_B 
\sin\theta \cos\theta \cos\chi \sin\theta_B
(H_{-\frac{1}{2}0}H_{\frac{1}{2}1} - H_{\frac{1}{2}0}H_{-\frac{1}{2}-1})
\bigg\}\bigg]\,\, . 
\end{align}
}
 
We have performed several checks on the correctness of the signs of the 
azimuthal correlation terms by using the semi--covariant representation 
Eq.~(\ref{lorentztrick})
and even doing a full-fledged covariant calculation\footnote{It is 
fair to say that the transcription of the results of a covariant calculation into the
helicity frame results used in this paper takes considerable amount of calculational 
effort.}. 
The overall sign of the nonflip azimuthal correlation terms (sixth and seventh line in 
Eq.~(\ref{joint2})) 
corrects the sign mistake in \cite{Korner:1991ph}. Note the reciprocity of the angular
decay distributions Eq.~(\ref{lside2}) and Eq.~(\ref{joint2}). One obtains Eq.~(\ref{joint2})
from Eq.~(\ref{lside2}) by the substitutions $(1+sign\{\lambda_2-\lambda_W\}P\cos\theta_P
\to (1+sign\{\lambda_2\}\alpha_B\cos\theta_B) $ for the polar correlation terms and
$P \sin\theta_P H_{\lambda_2\lambda_W} H_{\lambda_2\lambda'_W} \to
\alpha_B \sin\theta_B H_{\lambda_2\lambda_W} H_{-\lambda_2\lambda'_W}$ in the azimuthal
correlation terms.

Eq.~(\ref{joint2}) can be cast into a form where the dependence on the polarization
vector of the daughter baryon becomes explicit. One has
\begin{align}
\label{joint3}
\frac{d\Gamma}{dq^2 d\cos\theta d\chi d\cos\theta_B}=&
B(B_2 \to B_3 + \pi)\frac{1}{4}\frac{G_F^2}{(2\pi)^4}|V_{us}|^2
\frac{(q^2-m_l^2)p}{8M_1^2q^2}L_{\mu\nu}H^{\mu\nu}\times\nonumber \\ 
&\times\left(1+P_z\alpha_B\cos\theta_B + P_x\alpha_B \cos(\pi-\chi) \sin\theta_B\right)\, ,
\end{align} 
where $L_{\mu\nu}H^{\mu\nu}$, $P_z$ and $P_x$ are given in Eqs.~(\ref{lepton-hadron1}),
(\ref{polz}) and (\ref{polx}), respectively. When integrating Eq.~(\ref{joint3}) over $\cos\theta$,
$\cos\theta_B$ and $q^2$ one can define a mean azimuthal
correlation parameter $\langle\gamma\,\rangle$ through the relation 
$\Gamma\sim 1 + \langle\gamma\,\rangle\cos\chi$.
Using again the minimal form factor model of Sec.~\ref{section-helicity-ampt} 
one finds the numerical values 
$\langle\gamma\,\rangle=-0.44$ and $\langle\gamma\,\rangle=-0.13$ in the $e$-- and 
$\mu$--modes, respectively, for the mean azimuthal correlation parameter.

At zero--recoil one finds a rather simple expression for the above azimuthal correlation 
parameter. It reads
\begin{equation}
\label{gamma}
\gamma_{\rm zero\,recoil}=\frac{\alpha_B \pi^2}{16}\frac{1-2\sqrt{2}\frac{m_l^2}{2q^2}
H_{\frac{1}{2}t}^V/H_{\frac{1}{2}1}^A}{1+\frac{m_l^2}{2q^2}(1+2
|H_{\frac{1}{2}t}^V|^2 / |H_{\frac{1}{2}1}^A|^2 )}\, .
\end{equation}
which gives $\gamma=-0.61$ and $\gamma=-0.17$, respectively, in the $e$-- and $\mu$--mode 
in the minimal form factor model described in Sec.~\ref{section-helicity-ampt}.
Eq.~(\ref{gamma}) shows that, in the $e$--mode and at zero 
recoil, the azimuthal correlation parameter is independent of the form 
factors
as stated before in \cite{Korner:1991ph}. In the $\mu$--mode, however, the azimuthal 
correlation parameter at zero recoil does depend on the form factors through the ratio 
$H_{\frac{1}{2}t}^V/H_{\frac{1}{2}1}^A$. Since 
$H_{\frac{1}{2}t}^V/H_{\frac{1}{2}1}^A = F_{1}^V/(\sqrt{2}F_1^A)$ at zero recoil
(assuming $F_{3}^{V}=F_{2}^{A}=0$) this would afford the opportunity to determine the 
ratio $F_{1}^V/F_1^A$ through a zero recoil or near zero recoil measurement 
of the azimuthal correlation parameter in the $\mu$--mode. 
\section{Monte Carlo event generation and sample plots \label{section-montecarlo}}

In this section we present a few sample distributions generated from our event generator
in order to demonstrate the viability of our generator. As dynamical input
for the form factors we used the minimal form factor model described at the end of 
Sec.~\ref{section-helicity-ampt}, or slight 
variations on it. Of course, any other dynamical model can be used as input in the
event generator. For the angular decay distribution we used the full five-fold decay 
distribution from Appendix~\ref{appendix-5fold} describing the full decay chain 
$\Xi^0(\uparrow) \to \Sigma^+(\to p +\pi^0) + W^-_{\rm{off-shell}}
(\to l^- + \bar{\nu}_l)$. Masses and the decay asymmetry parameter are taken from
\cite{pdg08}. 

\begin{figure}[htbp]
  \begin{center}
    \epsfig{figure=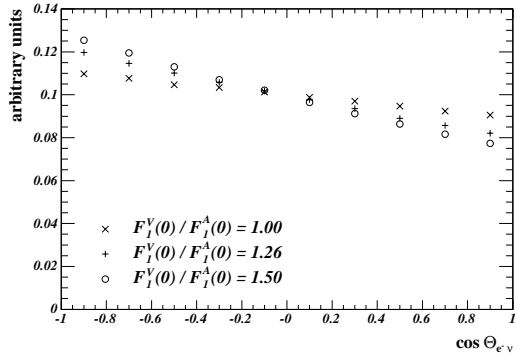, width=7.5cm}
     \caption{\label{g1f1_variation} \small \sf
Angle between the electron and the neutrino in the 
$\Xi^0$ rest frame for different values of the ratio $F_1^A(0)/F_1^V(0)$.}
  \end{center}
\end{figure} 
\begin{figure}[htbp]
  \begin{center}
    \epsfig{figure=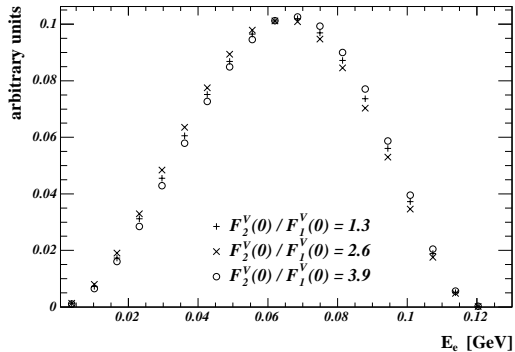, width=7.5cm}
     \caption{\label{f2f1_variation} \small \sf
Energy spectrum of the electron for different 
values of the ratio $F_2^V(0)/F_1^V(0)$ in the $\Xi^0$ rest frame.}
  \end{center}
\end{figure}
\begin{figure}[htbp]
  \begin{center}
    \epsfig{figure=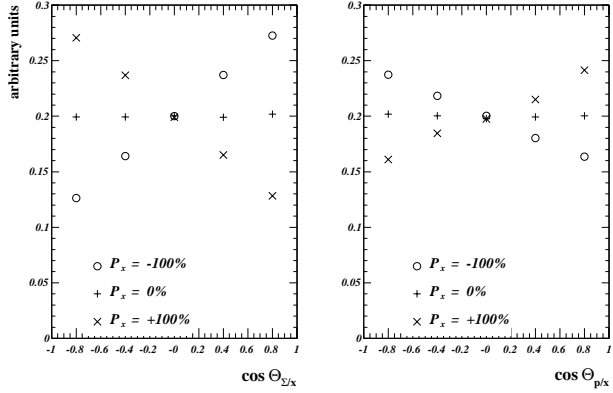, width=8.5cm}
     \caption{\label{pol_variation} \small \sf
Left: Angle between $\Sigma^+$ and $x$ axis in the 
$\Xi^{0}$ rest frame. Right: Angle between proton and $x$ axis in the $\Sigma^+$ rest 
frame for different initial state polarizations of the $\Xi^0$ hyperon.}
  \end{center}
\end{figure}

\begin{figure}[ht!]
  \begin{center}
    \epsfig{figure=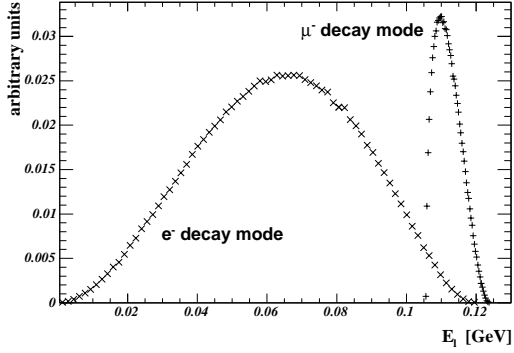, width=7.5cm}
     \caption{\label{lepton_spectra} \small \sf
Comparison of lepton spectra for the 
e-- and $\mu$--mode}
  \end{center}
\end{figure}

The implementation was done as follows. We first generated the 3-body phase space of
the primary decay $\Xi^0 \to \Sigma^+ + W^-_{\rm{off-shell}}$ using the widely used 
function {\em genbod} from the CERNLIB \cite{CERNLIB} library. Without loss of generality,
the axis of the initial state polarization of the parent baryon was chosen to point 
along the lab $x$--axis. The momenta of the decay products of the secondary 
decays $\Sigma^+\to p +\pi^0$ and $W^-_{\rm{off-shell}}\to l^- + \bar{\nu}_l$ were 
generated with uniformly distributed directions. Since the secondary decays are  
two-body decays, the moduli of the respective momenta are fixed. The resulting momentum 
vectors were used to obtain the angles and momenta needed to calculate the value of the 
matrix element. This result was multiplied by the phase space factor of the primary 
decay returned by the {\em genbod} routine. Applying an acceptance-rejection method, the 
whole procedure was repeated until a generated event was no longer discarded.

In Fig.~\ref{g1f1_variation} we show a plot of the dependence of the rate on the
angle between the electron and the neutrino in the $\Xi^0$ rest frame. In order to
exhibit the sensitivity of this distribution to the form factor ratio 
$F_1^A(0)/F_1^V(0)$ we also show corresponding distributions with slightly varied 
form factor ratios. We mention that the distribution Fig.~\ref{g1f1_variation} and
the following distributions are normalized to unity.

In Fig.~\ref{f2f1_variation} we show the electron energy spectrum
and its dependence on the form factor ratio $F_2^V(0)/F_1^V(0)$.
In Fig.~\ref{pol_variation} we show plots of the angular dependence of the angle
between the $\Sigma^+$ and the $x$ axis in the $\Xi^{0}$ rest frame (left) and between 
the proton and the $x$ axis in the $\Sigma^+$ rest frame (right) for different 
polarizations of the parent $\Xi^0$ hyperon. In order to demonstrate the dependence
on lepton mass effects, in Fig.~\ref{lepton_spectra} we show the energy spectra of the 
electron and the muon in the rest frame of the parent $\Xi^0$.

\section{Summary and conclusions \label{section-summary}}
We have worked out the angular decay distributions that govern the
semileptonic cascade decay 
$\Xi^0 \to \Sigma^+(\to p +\pi^0) + W^-_{\rm{off-shell}}(\to l^- + \bar{\nu}_l)$
using a cascade-type analysis. The cascade-type analysis has certain advantages,
the main advantage being that one obtains the decay distributions in a compact 
quasi-factorized form. This leads to rather compact forms for the decay
distributions. In our analysis we have included lepton mass
effects as well as polarization effects of the decaying parent hyperon. We have
always indicated the necessary sign changes when going from the $(l^-,\bar{\nu}_l)$ 
case to the $(l^+,\nu_l)$ case. Our angular decay formulae are thus applicable
also to the semileptonic hyperon decay $\Sigma^+ \to \Lambda + e^+ + \nu_e$, or to
semileptonic charm baryon decays induced by the transition $c \to s + l^+ +\nu_l$
and also to the decays $t \to b + l^+ +\nu_l$ .
It should be clear that our angular decay formula are also applicable to the corresponding
nonleptonic baryon decays involving vector mesons $(\lambda_W=\pm,0)$ or
pseudoscalar mesons $(\lambda_W=t)$. In this case one has to omit the interference
contributions between the time-component and the space-components of the currents.

Of interest are also the corresponding semileptonic antihyperon decays. The angular 
decay distributions of semileptonic antihyperon decays can be obtained from the
corresponding angular decay distributions of the semileptonic hyperon decays
by the replacements 
$H_{\lambda_2 \lambda_W}(B) \to H_{\lambda_2 \lambda_W}(\bar{B})$, 
$\alpha_B \to \alpha_{\bar{B}}$ and changing from the $(l^-,\bar{\nu}_l)$ to the 
$(l^+,\nu_l)$ case (or vice versa). Neglecting again $CP$--violating effects one
has from $CP$-invariance  
$H_{\lambda_2, \lambda_W}(\bar{B}) = H_{-\lambda_2, -\lambda_W}(B)$ and
$\alpha_{\bar{B}} =-\alpha_B$. One can verify that the decay distributions
in Eqs.~(\ref{lside2}), (\ref{hside2}) and (\ref{joint2}) are form invariant under 
$H_{\lambda_2 \lambda_W} \to  H_{-\lambda_2 -\lambda_W}$,
$\alpha_B  \to -\alpha_B$ and $P \to - P$ as follows from $CP$--invariance. 
We mention that the NA48 collaboration has recently observed the decay
$\overline{\Xi^{0}} \to \overline{\Sigma^{+}}e^{+}\nu_{e}$ and, 
based on 555 events, have given a branching ratio of 
$(2.55\pm0.14_{{\rm stat}}\pm0.10_{{\rm syst}})
\cdot 10^{-4}$ for this semileptonic antihyperon decay \cite{Batley:2006fc}.

We have summed over the helicity states of the final particles assuming that their 
polarization go unobserved. This corresponds to taking the trace of the density
matrix of the final particles. It is clear that one can equally well calculate 
the density matrix of the final state particles by leaving the relevant helicity index 
unsummed. This was illustrated for the density matrix of the final lepton in the 
semileptonic decay process.

Doing the helicity sums in the master formulas listed in this paper by hand can become 
quite cumbersome. However, this task can be automated and can be left to the computer.
The relevant {\it Mathematica} codes can be obtained from A.~Kadeer.
We mention also that the helicity frame analysis used in this paper can be easily
transcribed to a transversality frame analysis (see e.g. \cite{ks90}) where the $z$--axis
is perpendicular to the hadron plane . In fact, any choice
of $z$--axis in the analysis will provide the same total amount of information on the 
dynamics of the process entailed in the invariant amplitudes. It is then a question
of experimental exigency of whether to analyze angular decay distributions in the
helicity frame or the transversality frame, or, for that matter, in any other frame.
  
For the sake of conciseness we have written our results in terms of bilinear
products of the helicity amplitudes 
$H_{\lambda_2 \lambda_W}=H_{\lambda_2 \lambda_W}^{V}+H_{\lambda_2 \lambda_W}^{A}$
instead of bilinear products of the vector and axial vector helicity amplitudes 
$H_{\lambda_2 \lambda_W}^{V}$ and $H_{\lambda_2 \lambda_W}^{A}$. Writing the decay
distributions in terms of $H_{\lambda_2 \lambda_W}^{V}$ and 
$H_{\lambda_2 \lambda_W}^{A}$ can be quite illuminating if one wishes to identify the 
overall parity nature of the observables that multiply the angular terms in the angular 
decay distributions. 

We have formulated a minimal form factor model assuming $SU(3)$ symmetry at zero
momentum transfer and a canonical $q^{2}$--dependence of the form factors. We have 
used this minimal form factor model to numerically calculate a few 
observables such as branching rates, the rate ratio $\Gamma(e)/\Gamma(\mu)$, a 
lepton--side forward-backward asymmetry, the longitudinal and transverse polarizations 
of the daughter baryon and the lepton and a mean azimuthal correlation parameter in the decay
$\Xi^0 \to \Sigma^+(\to p +\pi^0) + l^- + \bar{\nu}_l$.

We have written a Monte Carlo event generator which is based on the angular decay
distributions derived in this paper. Among others, the event generator allows one
to generate events in the parent baryon rest frame.
The MC program can be obtained from Rainer Wanke
of the NA48 Collaboration (Rainer.Wanke@uni-mainz.de). We have 
presented a few decay distributions and correllations based on this event generator. 
We have, however, not systematically
investigated which observables would be optimal to obtain the maximal possible 
information on the underlying dynamics encapsuled in the invariant form factors or the
helicity amplitudes. A discussion of these issues also using parent baryon rest 
frame decays can be found in e.g. \cite{gk85}.

We did not provide an in--depth analysis of the dynamics of semileptonic hyperon decays
as is necessary if one wants to extract a value of the CKM matrix element $V_{us}$
from semileptonic hyperon decay data. This issue was discussed in 
\cite{Flores-Mendieta:1998ii,Cabibbo:2003ea,Cabibbo:2003cu,Mateu:2005wi}. We mention
that there has been some recent progress in the dynamical description of strangeness
changing semileptonic hyperon transition form factors from the lattice community
\cite{guadagnoli07,sasaki06,lin07} and in the framework of chiral symmetry
\cite{faessler08,Ledwig:2008ku} including the explicit calculation of chiral corrections 
\cite{villadoro06,lacour07,Jiang:2008aqa}.

We finally emphasize that we have not included $CP$--violating effects or radiative 
corrections in our analysis. The latter can be included using the results of 
\cite{gk85}. It will be interesting to find out how the radiative corrections will 
affect the angular decay distributions.
\vskip 0.8in

\vspace{1cm} {\bf Acknowledgements:}
One of the authors (J.\,G.\,K.) acknowledges discussions, and e-mail and fax 
exchanges with I.~Shipsey on whose insistence the sign mistake of
the azimuthal correlation term in \cite{Korner:1991ph} was discovered and rectified. 
J.\,G.\,K. would also like to thank K.~Zalewski for sharing his many insights into the
subject of angular decay distributions in particle physics. U.M. would like to thank 
E.C.~Swallow for valuable discussions. A.~Kadeer acknowledges the support of the DFG 
(Germany) through the Graduiertenkolleg ``Eichtheorien'' at the University of Mainz.

\begin{appendix}

\section{Two-body decay of a polarized particle in the helicity formalism \label{appendix-2body-master}}
\setcounter{equation}{0}\def\theequation{A\arabic{equation}}
In deriving the two--body decay of a polarized spin $J$ particle in the helicity formalism 
we shall closely follow the approach of \cite{jackson66,tung85}.
Consider the two particle decay $a \to b + c$ of a spin $J_a$ particle where the 
polarization of particle
$a$ in the frame $(x_0,y_0,z_0)$ is given by $\rho^0_{\lambda_a\lambda'_a}$. Consider
a second frame $(x,y,z)$ obtained from $(x_0,y_0,z_0)$ by the rotation $R(\theta,\phi,0)$
and whose $z$--axis is defined by particle $b$. The polarization 
density matrix $\rho$ in the frame $(x,y,z)$ is obtained by a ``rotation'' of the density
matrix $\rho^0$ from the frame $(x_0,y_0,z_0)$ to the frame $(x,y,z)$. 
The rate for $a \to b + c$ is then given by the
the sum of the decay probabilities $|H_{\lambda_b\lambda_c}|^2$ (with 
$\lambda_a=\lambda_b-\lambda_c$) weighted by the diagonal terms of the density
matrix $\rho$ of particle a in the frame $(x,y,z)$. Thus we find
\begin{align}
\label{basic}
\Gamma_{a \to b + c}(\theta,\phi) \propto 
\sum_{\lambda_a,\lambda'_a,\lambda_b,\lambda_c} &
|H_{\lambda_b\lambda_c}|^2 
D^{J*}_{\lambda_a,\lambda_b-\lambda_c}(\theta,\phi)\,
 \rho^0_{\lambda_a,\lambda'_a}
D^{J}_{\lambda'_a,\lambda_b-\lambda_c}(\theta,\phi) \,
\end{align}
where
\begin{equation}
D^J_{m,m'}(\theta,\phi)= e^{-im\phi}d^J_{m\,m'}(\theta)\,\, .
\end{equation}

All the master formulas written down in this paper can be obtained by a repeated application
of the basic two-body formula Eq.~(\ref{basic}).

\section{Wigner's $d^J$--functions for J=1/2 and J=1 \label{appendix-Wigner-D}}
\setcounter{equation}{0}\def\theequation{B\arabic{equation}}
For definiteness we list explicit forms for the two $d^J$--functions of Wigner used in 
this paper. We use the convention of Rose \cite{rose57} which is also the convention
of \cite{pdg08} . One has

\begin{equation}
d^{1/2}_{mm'}(\theta)=\left(
\begin{array}{cc}
\cos\theta/2 & -\sin\theta/2 \\ \sin\theta/2 & \cos\theta/2
\end{array}
\right)
\end{equation}
for $J={\textstyle \frac{1}{2}} $ and 
\begin{equation}
 d^1_{mm'}(\theta)=\left(
\begin{array}{ccc}
\frac{1}{2}(1+\cos\theta) \,\,\, & -\frac{1}{\sqrt{2}}\sin\theta\,\,\,  &\frac{1}{2}(1-\cos\theta) \\
\frac{1}{\sqrt{2}}\sin\theta  & \cos\theta \ & -\frac{1}{\sqrt{2}}\sin\theta \\
\frac{1}{2}(1-\cos\theta)\,\,\,  & \frac{1}{\sqrt{2}}\sin\theta\,\,\,  &\frac{1}{2}(1+\cos\theta) 
\end{array}
\right)
\vspace{0.5cm}
\end{equation}
\noindent for $J=1$. 
The rows and columns are labeled in the order $(1/2,-1/2)$ and $(1,0,-1)$,
respectively.

\section{$T$--odd contributions \label{appendix-T-odd}}
\setcounter{equation}{0}\def\theequation{C\arabic{equation}}
In the main text we have assumed that the invariant form factors and thereby the helicity 
amplitudes are relatively real. If one allows for relative phases between the
helicity amplitudes one will obtain so--called $T$--odd contributions in the angular
decay distributions. They appear in the azimuthal correlation terms as can be seen by
the following example taken from the joint angular decay distribution in 
Sec.~\ref{section-joint}. One of 
the azimuthal correlation terms derives from the helicity configurations 
$(\lambda_2=-1/2,\lambda_W=0;\lambda'_2=1/2,\lambda'_W=1)$ and
$(\lambda_2=1/2,\lambda_W=1;\lambda'_2=-1/2,\lambda'_W=0)$. Picking out the relevant
terms in the master formula Eq.~(\ref{joint1}) one has
\begin{align}
& H_{\frac{1}{2}1}H^*_{-\frac{1}{2}0}\,e^{i(\pi-\chi)}+
H_{-\frac{1}{2}0}H^*_{\frac{1}{2}1}\,e^{-i(\pi-\chi)}=
-2\cos\!\chi \,\rm{Re}(H_{\frac{1}{2}1}H^*_{-\frac{1}{2}0})
-2\sin\!\chi \,\rm{Im}(H_{\frac{1}{2}1}H^*_{-\frac{1}{2}0})\,\, .
\end{align}
The $\cos\!\chi$ dependent term already appears in Eq.~(\ref{joint2}) whereas the
$\sin\!\chi$ dependent term has been dropped in Eq.~(\ref{joint2}) because of the
relative reality assumption for the helicity amplitudes. Adding the relevant
$\theta$ and $\theta_B$ dependent trigonometric functions in the above azimuthal
correlation term one has the two angle dependent $T$--odd terms 
($\sin\theta \sin\!\chi \sin\theta_B  \,\rm{Im}(H_{\frac{1}{2}1}H^*_{-\frac{1}{2}0}$)) and
($\cos\theta\sin\theta \sin\!\chi \sin\theta_B\,\rm{Im}(H_{\frac{1}{2}1}
H^*_{-\frac{1}{2}0}$))
proportional to $\sin\!\chi$. 

Next we rewrite the product of angular factors in terms of scalar and pseodoscalar
products using the momentum representations in the $(x,y,z)$--system (see Fig.~\ref{five-fold}).
For the normalized momenta one has $(\hat{p}^2=1)$
\begin{eqnarray} 
\hat{p}_{l^-}&=&(\sin\theta \cos\chi,\sin\theta \sin\chi,-\cos\theta)\,, \\
\hat{p}_W&=&(0,0,-1) \,, \nonumber \\
\hat{p}_{\Sigma^+}&=&(0,0,1)\,,  \nonumber \\
\hat{p}_p &=& (\sin\theta_B,0,\cos\theta_B) \, ,\nonumber
\end{eqnarray} 
where the momenta have unit length indicated by a hat notation. The above T--odd 
angular factors can then be rewritten as
\begin{eqnarray} 
\label{t-odd}
\sin\theta \sin\chi \sin\theta_B &=&  \hat{p}_W\!\cdot\left(\hat{p}_{l^-}\!\times
\hat{p}_p\right)\,, \\
\cos\theta\sin\theta \sin\chi \sin\theta_B &=& (\hat{p}_{l^-}\cdot\hat{p}_W)\,  
[\, \hat{p}_W\cdot\left(\,\hat{p}_{l^-}\!\times\hat{p}_p\right)]\,. \nonumber
\end{eqnarray}  
Under time reversal ($t\to -t$) one has $(p\to-p)$. Since the $T$--odd momenta
invariants in Eq.~(\ref{t-odd}) involve an odd number of momenta they change sign 
under time reversal. This has led to the notion of the so--called $T$--odd obsevables. 
Observables that multiply $T$--odd momentum
invariants are called $T$--odd obsevables. They can be contributed to by true
$CP$--violating effects or by final state interaction effects unless either or both
change all helicity amplitudes by the same common phase. One may distinguish between
the two sources of $T$--odd effects by comparing with the corresponding antihyperon 
decays since phases from $CP$--violating effects change sign whereas phases from final 
state interaction effects do not change sign when going from hyperon to
antihyperon decays.

From the above example it should be clear how to obtain the $T$--odd contributions 
from the master formulas for the other cases. In practise what one has to do is
to add terms where the real part of the bilinear forms of helicity amplitudes is replaced
by the corresponding imaginary part and the cosine of the azimuthal angle is replaced by 
the sine with a possible sign change.
\section{Full five-fold angular decay distribution \label{appendix-5fold}}
\setcounter{equation}{0}\def\theequation{D\arabic{equation}}

\begin{figure}[t]
\begin{center}
\epsfig{figure=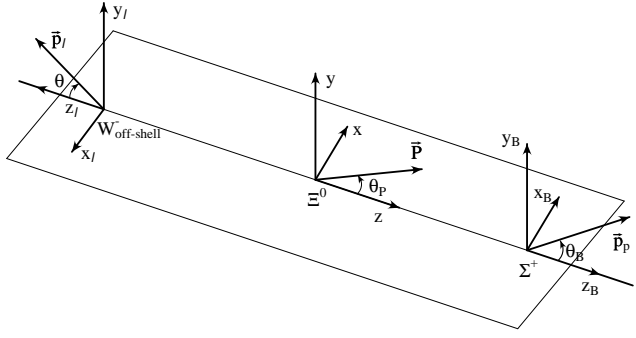, width=8.5cm}
\caption{\label{pol-angles} \small \sf
Definition of the three polar angles
$\theta$, $\theta_B$ and $\theta_P$ in the semileptonic decay of a 
polarized $\Xi^0$ into $ \Sigma^+ + l^- + \bar{\nu}_l$ followed by the 
nonleptonic
decay $\Sigma^+ \to p + \pi^0$. The polarization vector of the parent
baryon $\vec{P}$ lies
in the $(x,z)$--plane with positive $P_x$ component.}
\end{center}\end{figure}

\begin{figure}[t]\begin{center}
\epsfig{figure=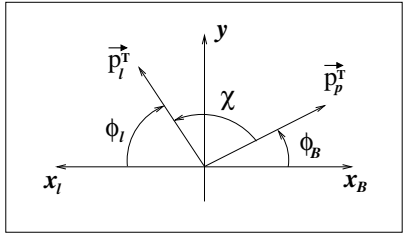, width=5.3cm}
\caption{\label{azi-angles}\small \sf
Definition of the three azimuthal angles
$\phi_l$, $\phi_B$ and $\chi$ ($\phi_l + \phi_B + \chi = \pi $) 
in the semileptonic decay of a 
polarized $\Xi^0$. Fig.~\ref{azi-angles} is a view of Fig.~\ref{pol-angles} from the right 
along the
negative $z$--direction. $\vec{p}_l^T$ and $\vec{p}_p^T$ denote the transverse
components of the momentum of the lepton and proton, respectively.}
\end{center}\end{figure}

In this appendix we write down the full five-fold angular decay distribution
for the semileptonic cascade decay of a polarized hyperon. There are now altogether
three polar angles $\theta$, $\theta_B$ and $\theta_P$, where $\theta_P$ describes
the polar orientation of the polarization vector of the parent hyperon as shown in 
Fig.~\ref{pol-angles} (which is directly taken from \cite{bkkz93}). Since 
there are now two planes in the cascade decay, there is one more
azimuthal angle which we choose as $\phi_l$ as shown in Fig.~\ref{azi-angles}. 
It is important to note that Fig.~\ref{azi-angles} shows a special
configuration where the momentum of the proton lies in the first quadrant and the 
momentum of the lepton lies in the second quadrant. It is clear that, for this special
configuration, the 
three azimuthal angles $\phi_l,\phi_B$ and $\chi$ add up to $\pi$ 
($\phi_l+\phi_B+ \chi= \pi$). 
For other configurations it may happen that the three angles add up to $\pi+ \rm{mod}(2\pi)$ 
if the rotation sense of the angles in Fig.~\ref{azi-angles} is kept. This 
will be of no consequence for the angular decay distribution which is invariant
under azimuthal $2\pi$ shifts. 

The full five-fold angular decay distribution can be directly taken from
\cite{bkkz93} after including the appropriate sign changes going from the
$(l^+,\nu_l)$ to the $(l^-,\bar{\nu}_l)$ case\footnote{Apart from listing angular decay
distributions Ref.~\cite{bkkz93} contains much additional useful material like e.g. a 
discussion of the statistical tensors of the processes and their bounds,
HQET results for heavy baryon transition form factors, etc..}.
We have simplified the corresponding 
expressions in \cite{bkkz93} by assuming as before that the helicity amplitudes are real.
For completeness we shall also write down the decay distribution in explicit
form using Wigner's $d^J$--functions as before. One has the master formula
\begin{align}
\label{five1}
W(\theta, \theta_P,\theta_B,\phi_B,\phi_l)  
\propto & \sum_{\lambda_l,\lambda_W,\lambda'_W,J,J',\lambda_2,
\lambda_2',\lambda_3} (-1)^{J+J'} |h^l_{\lambda_l\lambda_{\nu=\pm 1/2}}|^2
e^{i(\lambda_W-\lambda'_W)\phi_l}\times\nonumber \\ \nonumber
& \times\rho_{\lambda_2-\lambda_W,\lambda'_2-\lambda'_W}(\theta_{P})
d^J_{\lambda_W,\lambda_l-\lambda_{\nu}}(\theta)d^{J'}_{\lambda'_W,\lambda_l-\lambda_{\nu}}
(\theta)
H_{\lambda_2\lambda_W}H^*_{\lambda'_2\lambda'_W} \times\\
& \times e^{i(\lambda_2-\lambda'_2)\phi_B}d^{\frac{1}{2}}_{\lambda_2\lambda_3}(\theta_B)
d^{\frac{1}{2}}_{\lambda_2'\lambda_3}(\theta_B)
|h^B_{\lambda_3 0}|^2\,.
\end{align}

For the normalized five-fold angular decay distribution one finds

\begin{align}
\label{five2}
 \frac{d\Gamma}{dq^2 d\cos\theta_B d\cos\theta d\cos\theta_P d\chi d\phi_l} &=
B(B_2 \to B_3 + M) \frac{1}{12}\frac{G_F^2}{(2\pi)^5}|V_{us}|^2
\frac{(q^2-m_l^2)^2p}{8M_1^2q^2} \times \nonumber  \\ 
&\times \bigg[ b_{00}^{00}+3 \cos\theta \ b_{00}^{01}
+ \cos{\theta_{B}} \ b_{00}^{10} \nonumber  \\
& +(\frac{3}{2} \cos^{2}{\theta}-\frac{1}{2})  \ b_{00}^{02}\nonumber \\
& -3\sqrt{2} \sin{\theta}\cos{\phi_l} \ b_{01}^{01}     \nonumber \\
& -2         \sin{\theta_{B}}\cos{\phi_{B}} \ b_{10}^{10} \nonumber \\
& - \frac{3}{\sqrt{2}}\sin{2\theta}\cos{\phi_l} \ b_{01}^{02} \nonumber \\
& +3\cos{\theta}  \cos{\theta_{B}}\ b_{00}^{11}   \nonumber \\
& +(\frac{3}{2} \cos^{2}{\theta}-\frac{1}{2}) \cos{\theta_{B}} 
                   \ b_{00}^{12} \nonumber \\
& -\frac{3}{2}\sqrt{2} \sin{\theta}\sin{\theta_{B}}\cos{\chi}
                      \ b_{11}^{11}    \nonumber \\
& - \frac{3}{4}\sqrt{2} \sin{\theta_{B}}\sin{2\theta}\cos{\chi}
                      \ b_{11}^{12}     \nonumber \\
& +\frac{3}{2} \sin^{2}\theta \sin{\theta_{B}} \cos (\chi-\phi_l) 
                 \ b_{12}^{12}     \nonumber \\
& -6 \cos{\theta} \sin{\theta_{B}} \ \cos \phi_{B} 
                 \ b_{10}^{11}       \nonumber \\
& -3\sqrt{2} \sin{\theta}\cos {\theta_{B}}\cos{\phi_l}
                      \ b_{01}^{11}        \nonumber \\
& -\frac{3}{2}\sqrt{2} \sin{2\theta}\cos {\theta_{B}}\cos{\phi_l}
                      \ b_{01}^{12}         \nonumber \\
& -\sin{\theta_{B}}(3 \cos^{2}{\theta}-1)
                 \cos \phi_{B} b_{10}^{12} \,\,\,\bigg]\,.
\end{align}
It is important that the rotation sense of the azimuthal angles in 
Fig.~\ref{azi-angles} is kept.
We have used the relation $\phi_l + \phi_B + \chi = \pi +\rm{mod}(2\pi)$ to
rewrite $\cos(\phi_B + \phi_l)=-\cos\chi$ and $\cos(\phi_B + 2\phi_l)=-\cos(\chi-\phi_l)$.
Note that Eq.~(\ref{five2}) contains the redundant angle $\phi_B$. As before 
one can reexpress $\cos\phi_B$ as $\cos\phi_B = -\cos(\phi_l + \chi)$.
    
The coefficients $b^{kl}_{ij}$ in Eq.~(\ref{five2}) are given by \footnote{The coefficient
$b_{10}^{11}$ takes twice the value as compared to the corresponding coefficient in
\cite{bkkz93}. Also in Eq.~(48) of \cite{bkkz93} concerning the overall normalization 
one has to effect the replacement $q^2 \to (q^2-m_l^2)^2 /q^2$.} 
{\allowdisplaybreaks
\begin{align}
b_{00}^{00}= &\Big( (1+\epsilon) (|H_{-\frac{1}{2}-1}|^2
+|H_{\frac{1}{2}0}|^2) +3\epsilon|H_{\frac{1}{2}t}|^2\Big)
\,\rho_{\frac{1}{2}\frac{1}{2}} \nonumber \\
& + \Big((1+\epsilon)(|H_{\frac{1}{2}1}|^2
+|H_{-\frac{1}{2}0}|^2) + 3\epsilon |H_{-\frac{1}{2}t}|^2 \Big)
\rho_{-\frac{1}{2}-\frac{1}{2}},  \nonumber \\
b_{00}^{01}=&\frac{1}{2} \Big( 
\mp |H_{\frac{1}{2}1}|^{2}-
4\epsilon \ H_{-\frac{1}{2}0}H_{-\frac{1}{2}t}\Big) \, \rho_{-\frac{1}{2}-\frac{1}{2}} 
\nonumber \\
&-\frac{1}{2} \Big( 
\mp |H_{-\frac{1}{2}-1}|^{2}+
4\epsilon \ H_{\frac{1}{2}0}H_{\frac{1}{2}t}   
                                 \Big)\, \rho_{\frac{1}{2}\frac{1}{2}}\,, \nonumber\\
b_{00}^{10}=&\alpha_{B}   \Big(
(1+\epsilon)(-|H_{-\frac{1}{2}-1}|^2+|H_{\frac{1}{2}0}|^2)
+3\epsilon|H_{\frac{1}{2}t}|^2       
\Big)\rho_{\frac{1}{2} \frac{1}{2}}\nonumber \\
&-\alpha_{B}   \Big( 
(1+\epsilon) (|H_{-\frac{1}{2}0}|^2 \!-\! 
|H_{\frac{1}{2}1}|^2) \!+\! 3\epsilon|H_{-\frac{1}{2}t}|^2                 
\Big)\rho_{-\frac{1}{2}-\frac{1}{2}}, \nonumber \\
b_{00}^{02}=&\frac{1-2\epsilon}{2} 
     \Big( -2|H_{-\frac{1}{2}0}|^{2}+|H_{\frac{1}{2}1}|^{2}\Big)\,\rho_{-\frac{1}{2}-\frac{1}{2}}
\nonumber \\
&+\frac{1-2\epsilon}{2} 
\Big( |H_{-\frac{1}{2}-1}|^{2}-2|H_{\frac{1}{2}0}|^{2}\Big)\,\rho_{\frac{1}{2}\frac{1}{2}} \,,\nonumber \\
b_{01}^{01}=&\frac{1}{2} \Big( 
2\epsilon H_{-\frac{1}{2}t}H_{-\frac{1}{2}-1} \pm
H_{-\frac{1}{2}0}H_{-\frac{1}{2}-1} \nonumber \\
&-2\epsilon H_{\frac{1}{2}1}H_{\frac{1}{2}t} \pm
H_{\frac{1}{2}1}H_{\frac{1}{2}0}
                   \Big) \, \rho_{\frac{1}{2}-\frac{1}{2}} \,,\nonumber \\
b_{10}^{10}=&-\alpha_{B}   \Big( 
3\epsilon
H_{\frac{1}{2}t}H_{-\frac{1}{2}t}+
(1+\epsilon)H_{\frac{1}{2}0}H_{-\frac{1}{2}0}
  \Big) \, \rho_{-\frac{1}{2}\frac{1}{2}} \,,\nonumber \\
b_{01}^{02}=&\frac{1-2\epsilon}{2} 
       \Big(
 H_{-\frac{1}{2}0}H_{-\frac{1}{2}-1}-H_{\frac{1}{2}1}H_{\frac{1}{2}0}
        \Big)\, \rho_{\frac{1}{2}-\frac{1}{2}} \,,\nonumber \\
b_{00}^{11}=&\frac{\alpha_{B} }{2} \Big( 
\mp |H_{\frac{1}{2}1}|^{2}+
4\epsilon \ H_{-\frac{1}{2}0}H_{-\frac{1}{2}t}\Big)   
    \, \rho_{-\frac{1}{2}-\frac{1}{2}} \nonumber \\
&+\frac{\alpha_{B} }{2} \Big( 
\mp |H_{-\frac{1}{2}-1}|^{2}-
4\epsilon \ H_{\frac{1}{2}0}H_{\frac{1}{2}t}\Big)   
                                  \, \rho_{\frac{1}{2}\frac{1}{2}}\,, \nonumber \\
b_{00}^{12}=&\frac{\alpha_{B} }{2}(1-2\epsilon) 
     \Big( 2|H_{-\frac{1}{2}0}|^{2}+|H_{\frac{1}{2}1}|^{2}\Big)\,\rho_{-\frac{1}{2}-\frac{1}{2}}
\nonumber \\
&-\frac{\alpha_{B} }{2}(1-2\epsilon) 
 \Big( |H_{-\frac{1}{2}-1}|^{2}+2|H_{\frac{1}{2}0}|^{2}\Big)\,\rho_{\frac{1}{2}\frac{1}{2}} \,,\nonumber \\
b_{11}^{11}=&\alpha_{B} 
     \Big( 
2\epsilon H_{\frac{1}{2}1}H_{-\frac{1}{2}t} \mp
H_{\frac{1}{2}1}H_{-\frac{1}{2}0}
      \Big)\,\rho_{-\frac{1}{2}-\frac{1}{2}} \nonumber \\
& +\alpha_{B}   
     \Big(  -2\epsilon H_{\frac{1}{2}t}H_{-\frac{1}{2}-1} \mp
H_{\frac{1}{2}0}H_{-\frac{1}{2}-1}
        \Big)\, \rho_{\frac{1}{2}\frac{1}{2}} \,,\nonumber \\
b_{11}^{12}=&\alpha_{B} (1-2\epsilon) 
       \Big(
 H_{\frac{1}{2}1}H_{-\frac{1}{2}0}\,\rho_{-\frac{1}{2}-\frac{1}{2}}
-H_{\frac{1}{2}0}H_{-\frac{1}{2}-1}\,\rho_{\frac{1}{2}\frac{1}{2}}
        \Big)  \,, \nonumber \\
b_{12}^{12}=&-\alpha_{B}(1-2\epsilon) 
   H_{\frac{1}{2}1}H_{-\frac{1}{2}-1}\,\rho_{\frac{1}{2}-\frac{1}{2}}\,, \nonumber \\
b_{10}^{11}=&\alpha_{B}\, \epsilon \, 
                 \Big( 
H_{\frac{1}{2}0}H_{-\frac{1}{2}t}+
H_{\frac{1}{2}t}H_{-\frac{1}{2}0}  
                  \Big) \,\rho_{-\frac{1}{2}\frac{1}{2}} \,, \nonumber\\
b_{01}^{11}=&\frac{\alpha_{B}  }{2} 
                  \Big( -2\epsilon H_{-\frac{1}{2}t}H_{-\frac{1}{2}-1} \mp
H_{-\frac{1}{2}0}H_{-\frac{1}{2}-1}  \nonumber \\
&-2\epsilon H_{\frac{1}{2}1}H_{\frac{1}{2}t} \pm
H_{\frac{1}{2}1}H_{\frac{1}{2}0}
                   \Big) \, \rho_{\frac{1}{2}-\frac{1}{2}}\,, \nonumber \\
b_{01}^{12}=&-\frac{\alpha_{B} }{2}(1-2\epsilon) 
       \Big(
 H_{-\frac{1}{2}0}H_{-\frac{1}{2}-1}+H_{\frac{1}{2}1}H_{\frac{1}{2}0}
        \Big)\, \rho_{\frac{1}{2}-\frac{1}{2}}\,,\nonumber \\
b_{10}^{12}=& \alpha_{B} (1-2\epsilon) 
     H_{\frac{1}{2}0}H_{-\frac{1}{2}0}\, \rho_{-\frac{1}{2}\frac{1}{2}}\,.
\end{align}
}

We have introduced the abbreviation $\epsilon=m_l^2/2q^2$ for the leptonic flip
suppression factor. As in the main text the 
upper signs in the coefficients $b^{kl}_{ij}$ hold for the case 
$(l^-, \bar{\nu}_l)$ relevant to the cascade decay 
$\Xi^0 \to \Sigma^+ ( \to p + \pi^0) + l^- + \bar{\nu}_l$ treated in
this paper. The lower signs hold for the case $(l^+, \nu_l)$ discussed
in \cite{bkkz93}. 
Finally, $\rho_{\lambda_{1}\lambda_{1}^{'}}$ is the spin density matrix of the
parent hyperon given in Eq.~(\ref{densitym}).

We have performed various checks on Eq.~(\ref{five2}). First we found it to
agree with the angular decay distribution derived from the master formula
Eq.~(\ref{five1}). We further checked that Eq.~(\ref{five2}) reduces to the 
decay distributions listed in the main text after integration or after setting
the relevant parameters to zero. We thus checked that
Eq.~(\ref{five2}) reduces to Eq.~(\ref{joint2}) when setting $P=0$. There is
a factor of $4\pi$ from the integration over $\cos\theta_P$ and $\phi_l$.  
Further Eq.~(\ref{five2}) reduces to Eq.~(\ref{lside2}) when setting
$\alpha_B=0$, dropping the branching ratio factor $B(B_2 \to B_3 + M)$ and replacing
$\phi_l$ by $(\pi-\chi)$. Also there is a factor $4\pi$ from the integration over
$\cos\theta_B$ and $\phi_B$. Finally, Eq.~(\ref{five2}) reduces to Eq.~(\ref{hside2}) when
integrating over $\phi_l$ and $\cos \theta$. As mentioned before we have assumed
that the helicity amplitudes (or the invariant amplitudes) are relatively real. 
Nonzero relative phases between the helicity amplitudes could arise from final state
interaction effects or from extensions of the SM that bring in $CP$--violating phases 
(see e.g. \cite{Korner:1990yx,Korner:1992kj}). In such a case one would have to keep 
the full phase structure contained in the master formula Eq.~(\ref{five1}) or in the 
original version of Eq.~(\ref{five2}) listed in \cite{bkkz93}. 

\end{appendix}

\end{document}